\documentclass{article}
\usepackage{graphicx}
\usepackage{fancyhdr}
\usepackage{amsmath}
\usepackage{amssymb}
\usepackage{amsfonts}
\usepackage[colorinlistoftodos]{todonotes}
\usepackage{bm}
\usepackage{slashed}
\usepackage{scalerel,stackengine}
\usepackage[round]{natbib}

\newcommand{\RHheaderline}{TTK-19-04, ELHC\_2018-005\\ January 2019}
\fancypagestyle{firstpage}
{
  
  \fancyhead[R]{\RHheaderline}
  \setlength{\headheight}{53pt}
}

\begin{document}

\title{Naturalness, Wilsonian Renormalization, and ``Fundamental Parameters" in Quantum Field Theory }
\author{Joshua Rosaler and Robert Harlander\\[2em]
\textit{Institute for Theoretical Particle Physics and Cosmology,}\\
\textit{RWTH Aachen University, D-52056 Aachen, Germany} }
\date{}

\maketitle


\thispagestyle{firstpage}

\begin{abstract}
The Higgs naturalness principle served as the basis for the so far failed prediction that signatures of physics beyond the Standard Model (SM) would be discovered at the LHC. One influential formulation of the principle, which prohibits fine tuning of bare Standard Model (SM) parameters, rests on the assumption that a particular set of values for these parameters constitute the ``fundamental parameters" of the theory, and serve to mathematically define the theory. On the other hand, an old argument by Wetterich suggests that fine tuning of bare parameters merely reflects an arbitrary, inconvenient choice of expansion parameters and that the choice of parameters in an EFT is therefore arbitrary. We argue that these two interpretations of Higgs fine tuning reflect distinct ways of formulating and interpreting effective field theories (EFTs) within the Wilsonian framework: the first takes an EFT to be defined by a single set of physical, fundamental bare parameters, while the second takes a Wilsonian EFT to be defined instead by a whole Wilsonian renormalization group (RG) trajectory, associated with a one-parameter class of physically equivalent parametrizations. From this latter perspective, no single parametrization constitutes the physically correct, fundamental parametrization of the theory, and the delicate cancellation between bare Higgs mass and quantum corrections appears as an eliminable artifact of the arbitrary, unphysical reference scale with respect to which the physical amplitudes of the theory are parametrized. While the notion of fundamental parameters is well motivated in the context of condensed matter field theory, we explain why it may be superfluous in the context of high energy physics. 
\end{abstract}

\newpage
\tableofcontents

\section{Introduction}

The naturalness principle has served as an extremely influential guide to model building in high-energy physics for the past several decades, notably functioning as the basis for theoretical arguments in favor of supersymmetry, and for the expectation of discovering signatures of physics Beyond the Standard Model (BSM) at the Large Hadron Collider (LHC). The principle has been formulated variously as a prohibition against ``unlikely" fine-tuning of Standard Model (SM) parameters, a prohibition on delicate sensitivity between physics at different energy scales, and as the requirement that all dimensionless parameters in an effective field theory be of order one unless protected by a symmetry.

The continuing absence of BSM physics up to ever higher energies, combined with the notorious quadratic dependence of the physical Higgs mass on heavy scales, entails increasingly strong violations of naturalness in all of these formulations. There remains an important open question concerning what to make of this fact. One possibility is to regard failure of naturalness as ``crying out for explanation" by some deeper theory. 
	\footnote{This characterization of the naturalness problem, in which the Higgs mass is said in some sense to ``cry out" or ``scream" for explanation, has been recently examined at length in \citep{hossenfelder2018screams}.
	}
	Another is to simply accept failure of naturalness as a brute fact. Still another is to abandon the Higgs naturalness principle entirely.

In favor of the first view, that failure of naturalness in the SM remains a special puzzle to be resolved by BSM theories, Giudice writes, ``If the LHC rules out dynamical solutions to Higgs naturalness at the weak scale, it does not eradicate the problem: a doctor who is unable to find the right diagnosis cannot simply declare the patient healed. Even in post-natural times, the concept of naturalness cannot simply be ignored ... One way or another, naturalness will still play a role in the post-naturalness era"  \citep{giudice2017dawn}. Thus, on Giudice's view, although we are entering a ``post-naturalness" era, failure of Higgs naturalness remains as much a problem as ever. On the other hand, several voices have urged abandonment of the naturalness principle itself - at least, as it has been applied in the cases of the Higgs and of the cosmological constant, which are closely analogous \citep{woitEdge},  \citep{hossenfelder2018screams}. Since our discussion below inclines toward this latter point of view, we wish to acknowledge at the outset that from one perspective, it may seem all too easy, now that failure of naturalness in the Standard Model has been increasingly empirically confirmed, to join the chorus of voices decrying naturalness. However, we emphasize that the status of the naturalness principle itself is far from settled by existing evidence: violation of naturalness by the Standard Model 
\footnote{As is sometimes noted, there is no sharp division between ``natural" and ``unnatural" theories; naturalness is a matter of degree, often associated with the degree of fine tuning or sensitivity to a theory's ``fundamental parameters." Thus, it is to some degree a matter of taste at  precisely what point one takes the SM to have violated the naturalness criterion. 
} 
does not in and of itself imply that the fault lies with the naturalness principle since it may be the case, as Giudice suggests, that failure of naturalness in the SM is a puzzle to be resolved by some BSM theory. 

Nevertheless, the naturalness principle's so far failed prediction of new physics at the LHC motivates careful re-examination of the arguments that have been advanced in its favor. 
Here, we probe a possible point of weakness in the rationale for one influential formulation of the naturalness principle, which precludes ``unlikely" or ``contrived" tuning of the Standard Model's bare parameters.
\footnote{In this discussion, we do not explicitly address alternative formulations of the naturalness criterion, such as those that preclude delicate sensitivity between physics at different scales, and those that require dimensionless parameters to be of order one. For extended philosophical analysis of naturalness in the first sense, see Williams' \citep{williams2015naturalness}; for a physicist's perspective on naturalness in this sense, see Giudice's \citep{giudice2013naturalness}. For discussion of naturalness in the second sense, see for example the original article by 't Hooft, \citep{hooft1980naturalness}, and Wells' \citep{wells2015utility}. 
}
We show how this rationale, advanced by some of the original advocates of the naturalness principle, is grounded in the notion that there exists a single physically correct, ``fundamental" parametrization for a given Wilsonian effective field theory (EFTs) in particle physics, such as the Standard Model or quantum electrodynamics. We then revisit an alternative but relatively little-discussed argument suggesting that fine tuning of bare parameters is unproblematic because it merely reflects an inconvenient choice of expansion parameters rather than an unlikely coincidence in need of explanation. We emphasize that this view of fine tuning appears implicitly to do without the notion of a fundamental parametrization, and thus to reflect an alternative understanding of the Wilsonian approach to effective field theory. In an effort to flesh out the details of this alternative view more explicitly, we sketch one possible interpretation of Wilsonian EFTs and the Wilsonian renormalization group (RG), according to which no single parametrization of the theory counts as more fundamental than any other, or as the ``true" parametrization of the theory. In this view, a Wilsonian EFT is not specified by any single parametrization, but instead by an entire Wilsonian RG trajectory, which specifies a one-parameter class of physically equivalent parametrizations. Since the mathematical apparatus of Wilsonian renormalization permits one to calculate correlation functions, and any observables constructed from them, using any one of this continuous infinity of parametrizations, we argue that the notion of fundamental parameters may constitute an idle metaphysical supposition. This view reflects a different formulation and physical interpretation of Wilsonian EFTs from the one based on fundamental parameters, in that it differs on the question of how a Wilsonian EFT is mathematically defined and on which parts of the theory's formalism represent real physical features in the world. However, we also underscore several respects in which this approach requires further development before it can be said to constitute a mathematically rigorous definition of quantum field theory. 
	
	In our discussion, we emphasize that one way of motivating the concept of fundamental parameters in high energy physics (HEP) has been by analogies with condensed matter physics (CMP), where the notion of a single fundamental set of parameters associated with the physical cutoff scale of the theory is more obviously appropriate. Indeed, this analogy served as a major source of inspiration for Wilson in the development of his approach to renormalization. Employing D. Fraser's distinction between ``formal" and ``physical" analogies, we argue that the strategy of motivating naturalness and the concept of fundamental parameters by analogies with condensed matter theory rests on what may be an overly physical interpretation of this analogy. On the other hand, we emphasize that several formulations of naturalness invoking the concept of fundamental parameters, including Susskind's original 1979 paper on naturalness, make no reference to the high-energy/condensed-matter analogy, and that our discussion therefore addresses just one possible motivation for the notion of fundamental parameters in QFT. 	

Our discussion is outlined as follows. Section \ref{Foundations} establishes several foundational assumptions about the formal definition of Wilsonian effective field theories that will serve to ground our later discussion of naturalness. Section \ref{FineTuningNaturalness} introduces the formulation of the Higgs naturalness problem as arising from the need for a delicate cancellation between the bare Higgs mass and its quantum corrections, and introduces two conflicting views of this cancellation: the first view takes this cancellation as an instance of mysterious fine tuning, while the other sees it as an unphysical and unproblematic artifact of convention. Section \ref{FundParams} reviews the historical motivations for imposing naturalness cited by original proponents of the naturalness principle, underscoring the reliance of these formulations on the notion of fundamental parameters. We attempt to state explicitly the core features of the interpretation of Wilsonian EFTs based on fundamental parameters, and consider the status of Higgs fine tuning from this perspective. Section \ref{Auxiliary} sketches one interpretation of Wilsonian EFTs in the absence of fundamental parameters, exploring the notion that an EFT is specified by its whole Wilsonian RG trajectory, points along which correspond to physically equivalent parametrizations of a single model rather than parametrizations of distinct models. We offer several arguments in support of this alternative view, and identify points where it requires further development. We show that from this perspective, Higgs fine tuning is an eliminable artifact associated with one conventional choice of bare parametrization, rather than a physical coincidence urgently demanding explanation by deeper theories. Finally, we consider the implications of this view for debates about the physical interpretation of quantum field theory. In Section \ref{FraserAnalogies}, we review D. Fraser's distinction between ``formal" and ``physical" analogies, and specifically between ``formal" and ``physical" interpretations of the HEP/CMP analogy, which helps to characterize the difference between the two views of Wilsonian EFT's considered here. Section \ref{Conclusion} is the Conclusion.

\section{Foundational Assumptions} \label{Foundations}

How is a quantum field theory mathematically defined? This is a notoriously difficult and controversial question, with the difficulties arising in large part as a consequence of the infinite number of degrees of freedom described by the theory, and the resulting infinities that occur in perturbative expansions of QFT amplitudes. There exist multiple rival research programs, including the algebraic and constructive approaches to quantum field theory, that have attempted in different ways to place QFT on firm mathematical foundations. However, continuum approaches such as the algebraic approach are only known to work in contrived and unphysical settings - for example, in which the number of spacetime dimensions is less than four - and thus far have had difficulty incorporating any of the physically realistic QFT's, such as the Standard Model or quantum electrodynamics, used to describe the phenomenology observed in accelerators like the LHC. Largely for this reason, we consider here the alternative approach based on Wilson's effective field theory approach to renormalization and quantum field theory, which takes quantum field theories such as the Standard Model and quantum electrodynamics to be defined with a finite cutoff, and thereby avoids the problematic infinities that occur in axiomatic continuum approaches. 
\footnote{The virtues and shortcomings of such a cutoff-based approach, as contrasted with axiomatic, continuum-based approaches such as algebraic quantum field theory (AQFT) and constructive field theory, have been debated extensively by Wallace and Fraser in \citep{wallaceNaivete}, \citep{wallace2011taking}, \citep{fraser2009quantum}, \citep{fraser2011take}. Largely for the reasons articulated by Wallace - in particular, the difficulty in axiomatic approaches of making contact with the empirical success of the Standard Model -  we choose to ground our discussion in a Wilsonian cutoff-based rather than an axiomatic approach to the formulation of quantum field theory. For technical discussion of constructive approaches to defining QFT on the continuum, see for example  \citep{rivasseau2014perturbative}. For philosophical discussion of constructive QFT see Hancox-Li \citep{hancox2017solutions}.  
}

Beyond the infinities that occur in perturbative expansions, there is the closely related problem of how to define a quantum field theory outside the context of perturbation theory. 
\footnote{For philosophical analysis of approaches to the foundations of QFT based on perturbation theory,
 see, for example, the recent work of Miller and J. Fraser  \citep{miller2016mathematical}, \citep{fraser2017real}.  
 } 
Such problems are especially salient in the context of low-energy QCD calculations, where perturbation theory breaks down. To address both problems, the Wilsonian strategy of defining a QFT with a finite cutoff is often employed; see for example \citep{collins1984renormalization} and \citep{montvay1997quantum}. As is often emphasized, this strategy permits one to numerically define the amplitudes of a QFT in a way that is completely finite and does not rely on the assumption of small coupling; perturbative expansions are then understood as generating approximations to these non-perturbatively defined quantities. In our discussion of naturalness below, we likewise understand QFT amplitudes to be non-perturbatively defined in this manner. Since QFTs so defined are generally understood to be empirically valid only up to some finite physical cutoff scale $\Lambda_{phys}$, they are often designated as ``effective field theories" (EFT's). However, as we discuss in later sections, there may be more than one way to interpret the Wilsonian formalism, and to make use of this formalism in efforts to give a rigorous definition of QFT - one based on the assumption that there exists a set of ``fundamental parameters" that serve to uniquely define the EFT, and another that treats different finite parametrizations on equal footing.

\subsection{Wilsonian Path Integral Approach to Non-Perturbative Calculation of QFT Observables}

In practice, generating predictions from an EFT proceeds via the calculation of n-point correlation functions, which can be defined non-perturbatively via the imposition of a finite cutoff $\Lambda$ in the expression for the path integral. S-matrix elements, cross sections, pole masses, and other physical quantities can then be constructed from the correlation functions. The cutoff helps to ensure that all such quantities are mathematically well-defined, in the sense that it is possible to assign a definite numerical value to them. 

All correlation functions, and any observables built from them, can be calculated from the Feynman path integral, which is mathematically defined via the imposition of a cutoff regulator - e.g., a hard momentum cutoff, a smooth momentum cutoff, or a real space lattice -  and Wick rotation to Euclidean spacetime. Adopting a lattice spacing $a$ for concreteness, which imposes an effective momentum cutoff $\Lambda$ of order $\frac{2\pi}{a}$,
	the path integral $\mathcal{Z[}J]$ for a theory with Lagrangian $\mathcal{L}(\phi(x))$ in the presence of a classical external source field $J(x)$ is

\begin{align} \label{PathIntegral}
\mathcal{Z}[J] &= \int^{\Lambda} \mathcal{D} \phi \ e^{i \int d^{4} x \left[ \mathcal{L}(\phi(x))  \ + \ J(x) \phi(x) \right]} \\
&\equiv \int \Pi_{i=1}^{M} d \phi_{i} \  e^{i a^{4} \sum_{i=1}^{M} \left[  \mathcal{L}(\phi_i) + J_i \phi_i \right] } , \nonumber
\end{align}

\noindent where $\mathcal{L}(\phi(x))$ is the ``bare" Lagrangian of the theory, $M=(\lfloor L/a \rfloor)^{4}$, $a$ is the lattice spacing on a hypercubic 4-D lattice (note that the time dimension has now been discretized as well), $L$ is the length of each edge of the lattice, $J_i \equiv J(x_{i})$ is a background source field, $\phi_i \equiv \phi(x_{i})$, and $x_i \equiv (n_0 a, n_1 a, n_2 a, n_3 a)$ for $n_{\alpha} \in \mathbb{Z}$ and $-\frac{L}{2a} \leq n_{\alpha} \leq \frac{L}{2a}$, $\alpha = 0, 1, 2, 3$. The lengths $a$ and $L$ function, respectively, as a UV regulator associated with an upper momentum cutoff of $\Lambda = \frac{2 \pi}{a}$  and an IR regulator associated with lower momentum cutoff $\frac{2 \pi}{L}$. 
\footnote{For the purposes of our discussion, only the UV regulator $a$ will be relevant; the IR regulator $L$ will not play a role. 
}

In keeping with the Wilsonian approach, the bare Lagrangian $\mathcal{L}$ is taken to consist of all terms consistent with a given choice of fields $\phi$ and symmetries; in the case where $\phi$ is a scalar field and the theory possesses $\phi \rightarrow - \phi$ symmetry,

\begin{align} \label{LagrangianBare}
\mathcal{L} & = \sum_{n} g_{n} \mathcal{O}_{n} \nonumber \\
&= g_{1} \left(\partial_{\mu} \phi\right)^{2} + g_{2} \phi^{2} + g_{3} \phi^{4} + g_{4} \phi^{6}  + g_{5}(\partial_{\mu} \phi)^{2} \phi^{2} +  g_{6} \phi^{8} + ... \ , 
\end{align}

\noindent where the bare parameters $g_{i}$ can be any real numbers. Here, we use the term ``bare" to refer to any parametrization of the theory of the form given by the path integral (\ref{PathIntegral}) and Lagrangian (\ref{LagrangianBare}); that is, a ``bare" parameter on our usage is a parameter appearing in the Lagrangian of the path integral, whatever the chosen value of $\Lambda$. 
\footnote{We should acknowledge here a distinct use of the term ``bare" within the context of Wilsonian renormalization, in which it is only the path integral parametrization for some \textit{particular} choice of cutoff $\Lambda_{0}$, associated with the physical cutoff scale of the theory, that constitutes the ``bare" parametrization. Parametrizations related to this parametrization via Wilsonian RG transformations are designated on this usage as ``renormalized." This usage reflects the presence in statistical mechanical or condensed matter applications of a sharply defined physical cutoff and fundamental bare parameters associated with a real physical lattice. By contrast, the usage of ``bare" employed here, in which bare parameters absorb the $\Lambda$ dependence of $Z[J]$ without being attached to any particular value of $\Lambda$, is consistent with perturbative renormalization schemes, in which the bare parameters are also not uniquely attached to any single value of $\Lambda$. 
	}
The calculation of physical quantities such as pole masses and cross sections from $\mathcal{Z}[J]$ is mediated via the calculation of $n$-point Green's functions $G^{(n)}(x_1, ... , x_n)$, which encode the full dynamics of the QFT:

\begin{align} \label{GreenFcns}
G^{(n)}(x_1, ... , x_n) & \equiv \frac{\int^{\Lambda} \mathcal{D} \phi \ \phi(x_1) \ ...  \ \phi(x_n) \ e^{i \int d^{4} x  \sum_n  g_n \mathcal{O}_n  } }{\int^{\Lambda} \mathcal{D} \phi  \ e^{i \int d^{4} x \sum_n  g_n \mathcal{O}_n }} \\
& = \frac{ (-i)^{n}}{\mathcal{Z}[0]} \frac{\delta^n}{\delta J(x_1) \  ... \ \delta J(x_n)}\bigg|_{J=0} \mathcal{Z}[J] \nonumber.
\end{align}

\noindent For example, the physical or ``pole" mass $m_{p}$ is calculated from the bare parameters as the lowest-lying pole of the Fourier transform of the two-point function,

\begin{align}\label{KallenLehmann}
G^{(2)}(x_1, x_ 2) &=  \int^{\Lambda} \frac{d^4 p}{(2\pi)^4} e^{ip(x_1 - x_2)}\left[ \frac{i Z(g, \Lambda)}{p^{2} - m_{p}^{2}(g, \Lambda)} +  \int_{\sim 4 m_{p}^{2}}^{\Lambda} d \mu  \ \frac{\rho(\mu)}{p^{2} - \mu^{2}} \right],  
\end{align}

\noindent where $\rho(\mu)$ is the so-called spectral density function, $g \equiv (g_{1}, g_{2}, g_3, .... )$; the symbol $\sim$ in the lower limit $\sim 4 m_{p}^{2}$ of the second integral is intended to signal the possibility of bound states below the threshold energy-squared for multi-particle production $(2 m_p)^{2}$. The field renormalization $Z(g,\Lambda)$ is defined as $Z(g, \Lambda) \equiv | \langle \lambda(0) |\hat{\phi}(0) |\Omega \rangle|^{2}$, where $|\lambda(0) \rangle$ is a one-particle energy eigenstate of the full interacting theory with zero spatial momentum. The S-matrix element for $2 $ $\rightarrow$ $n-2$ particle scattering can be calculated from Green's functions via the LSZ reduction formula:

\begin{align}
S(p_{1}, ... , p_{n}; g, \Lambda) & =    \tilde{G}^{(n)}(p_{1}, ... ,p_{n}; g, \Lambda) \ \ \tilde{G}^{(2), -1}(p_{1}; g, \Lambda) ... \tilde{G}^{(2), -1}(p_{n}; g, \Lambda). 
\end{align}

\noindent This in turn can be used to infer the differential and total cross section of the process in question. Note that the above relations are non-perturbative, and that all quantities defined are finite and in principle calculable from first principles.

\subsection{The Wilsonian Renormalization Group}

The Wilsonian renormalization group (RG) is a set of transformations on the parameters $g$ and $\Lambda$ of $\mathcal{Z}[J]$, that leave the value of $\mathcal{Z}[J]$ exactly unchanged. From this it follows that the values of physical quantities such as pole masses and (mod-squared) S-matrix elements computed from $\mathcal{Z}[J]$ remain exactly unchanged under these transformations.
\footnote{The ``mod-squared" in parentheses is intended to signal that S-matrix elements possess an unphysical global phase, so are likely not themselves directly physical. However, squared S-matrix elements, which are used to compute the cross sections measured in accelerators, do not depend on this global phase. 
}
Thus, the Wilsonian RG determines the cutoff dependence $g(\Lambda)$ that must be ascribed to bare Lagrangian parameters $g$ in order that $\mathcal{Z}[J]$ remain invariant. The process of transforming from a bare parametrization $(g(\Lambda), \Lambda)$ at one cutoff scale to another bare parametrization $(g(\Lambda'), \Lambda')$ at a slightly lower cutoff scale is implemented by splitting the path integral into an integration over field modes $\phi_{\Lambda'}$ with momenta less than or equal to some lowered cutoff $\Lambda' = \Lambda - \delta \Lambda$, and an integration over field modes $\phi_{\delta \Lambda} = \phi_{\Lambda} - \phi_{\Lambda'}$ with momenta greater than $\Lambda'$ but less than $\Lambda$, and then explicitly performing the latter integral:

\begin{align}
\mathcal{Z}[0] &= \int \mathcal{D} \phi_{\Lambda} \ e^{i \int d^{4} x \ \mathcal{L}(\phi_{\Lambda}; g(\Lambda)) } \nonumber \\
&= \int \mathcal{D} \phi_{\Lambda'} \left( \int \mathcal{D} \phi_{\delta \Lambda}  e^{i \int d^{4} x \ \mathcal{L}(\phi_{\Lambda} + \phi_{\delta \Lambda}; g(\Lambda))}   \right) \nonumber \\
&= \int \mathcal{D} \phi_{\Lambda'} e^{i \int d^{4} x \ \mathcal{L}(\phi_{\Lambda'}; g(\Lambda')) }
\end{align}

\noindent where $e^{i \int d^{4} x \ \mathcal{L}(\phi_{\Lambda'};  g(\Lambda')) } \equiv \int \mathcal{D} \phi_{\delta \Lambda} e^{i \int d^{4} x \ \mathcal{L}(\phi_{\Lambda'} + \phi_{\delta \Lambda};
  g(\Lambda)) }$. 
\footnote{Here we have assumed that in the expressions $\mathcal{L}(\phi_{\Lambda'};  g(\Lambda'))$ and  $\mathcal{L}(\phi_{\Lambda}; g(\Lambda)) $, $\mathcal{L}(\phi_{\Lambda}; g)$ is the most general function of $\phi$ allowed by the symmetries of the theory; thus, changes to the Lagrangian induced by the Wilsonian RG flow are determined entirely by changes in the coefficients $g$ specified by the RG trajectory $g(\Lambda)$.
} 
The effect of integrating over the field modes $\phi_{\delta \Lambda}$ is to alter the values of coefficients of the Lagrangian, yielding a new bare Lagrangian $\mathcal{L}(\phi_{\Lambda'}; g(\Lambda'))$, which is a function of the fields with lowered cutoff $\Lambda'$ with altered bare coefficients $g(\Lambda')$. Iterating this procedure by successively integrating over infinitesimal momentum shells, one obtains the continuous functional dependence $g(\Lambda)$ of the infinite set of bare parameters on the adjustable cutoff parameter $\Lambda$, where
$g(\Lambda) = (g_{1}(\Lambda), g_{2}(\Lambda), g_3(\Lambda), .... )$. This functional dependence is determined by an infinite set of coupled first-order differential equations, 

\begin{align} \label{BetaFcn}
\Lambda \frac{d g_{i}(\Lambda)}{d \Lambda} = \beta_{i}(g(\Lambda), \Lambda)
\end{align}

\noindent and in initial conditions $g(\Lambda_{0}) \equiv (g_{1}(\Lambda_{0}), g_{2}(\Lambda_{0}), g_3(\Lambda_{0}), .... )$ specifying the values of all bare parameters at some particular scale $\Lambda_{0}$. The beta functions $\beta_i$ can in principle be calculated from the path integral exactly by solving the Wilson-Polchinski equation, which serves as the basis for one approach to the non-perturbative or ``exact" renormalization group \citep{polchinski1984renormalization}, \citep{rosten2012fundamentals}. 
\footnote{There is a second popular approach to non-perturbative renormalization, which we do not discuss here, based on variation of an IR rather than a UV cutoff, and describing the scale dependence of parameters in the so-called effective average action, which is governed by the Wetterich equation \citep{wetterich1991average}, \citep{wetterich1993exact}, \citep{berges2002non}, \citep{delamotte2012introduction}.  
}
Assuming that the bare couplings are sufficiently small, one may also calculate the functions $\beta_{i}$ perturbatively, Taylor expanding the exponential of the interaction term in the Lagrangian and then using Wick's Theorem (or its path integral equivalent) to perform the integration over the high-energy field modes $\phi_{\delta \Lambda}$, treating the low-energy modes $\phi_{\Lambda'}$ as external fields.
\footnote{See, e.g., \citep{srednicki2007quantum}, Ch. 29., or \citep{peskin1995introduction}, Ch. 12 for details of this perturbative analysis.  
 }

\section{The Higgs Fine Tuning Naturalness Problem} \label{FineTuningNaturalness}

The original and perhaps most cited formulation of the Higgs fine tuning naturalness problem 
\footnote{We note that there exists a separate formulation of the Higgs fine tuning that concerns the running $\overline{MS}$ Higgs mass rather than the bare Higgs mass. While the formulation considered here concerns the relationship between the Higgs bare mass (in a cutoff-based scheme) and the Higgs pole mass, the $\overline{MS}$ formulation concerns the relationship between the $\overline{MS}$ scalar mass in an EFT where fields much heavier than the scalar have been integrated out, and the $\overline{MS}$ scalar in an EFT where these fields occur explicitly in the Lagrangian; see, e.g., \citep{skiba2010tasi}. We defer detailed consideration of this formulation to future work. 
}
rests on the observation that the leading contribution to the perturbative one-loop expansion of the (squared) Higgs pole mass $m^{2}_{p}$ in terms of bare SM parameters gives,

\begin{align} \label{FineTuning}
m^{2}_{p} &= m_{0}^{2} + \delta m^{2} \nonumber \\
 &= m_{0}^{2} - \frac{y_{t}^{2}}{8 \pi^{2}} \Lambda_{SM}^{2} + ... \nonumber \\
& = \Lambda_{SM}^{2}(\tilde{m}_{0} - \frac{y_{t}^{2}}{8 \pi^{2}}) + \ldots 
\end{align}

\noindent where $y_{t}$ is the top quark Yukawa coupling, $\Lambda_{SM}$ the Standard Model's physical cutoff (that is, the scale at which it ceases to be empirically valid),
\footnote{At leading non-trivial order perturbation theory, one can take $y_{t}$ to be either a bare coupling or a renormalized coupling, since the corresponding expressions only differ at higher orders in perturbation theory. 
}
$m_{0}^{2}$ is the bare Higgs mass, and $\tilde{m}_{0}^{2} \equiv \frac{\tilde{m}_{0}^{2}}{\Lambda_{SM}^{2}}$ is the dimensionless bare Higgs mass in units of $\Lambda_{SM}$. 
\footnote{See, for example, \citep{martin2010supersymmetry}, for a formulation of the Higgs naturalness problem along these lines. 
}
The corrections $\delta m^{2}$ also receive smaller contributions from lighter quarks, which can be neglected in the approximation where only the dominant correction to the Higgs bare mass is considered. By contrast with all other particle masses in the Standard Model, whose quantum corrections depend logarithmically on the cutoff, the Higgs mass undergoes quantum corrections that depend quadratically on the cutoff. Measurements at the LHC have further determined that $m^{2}_{p} = (125 \text{ GeV})^{2}$, and that $(1 \times 10^{3} \text{ GeV})^{2} \lesssim \Lambda_{SM}^{2} \lesssim (10^{19} \text{ GeV})^{2}$. While the lower limit $(1 \times 10^{3} \text{ GeV})^{2}$ has been set on the basis of LHC measurements, the upper limit, equal to the Planck scale, is set by theoretical expectations regarding the scales at which quantum gravitational effects can no longer be ignored. Together, these facts imply that the bare Higgs mass $m_{0}^{2}$ must be ``fine tuned" in order to recover the measured value of the physical, pole mass of the Higgs. The minimal degree of fine tuning required to recover the measured pole mass $m^{2}_{p} \sim \mathcal{O}(10^{4})$ increases with the empirically established lower bound on $\Lambda_{SM}^{2}$. At the lower limit of the allowed range for $\Lambda_{SM}$, $\Lambda_{SM} = 1 \times 10^{3} \text{ GeV}$, the relation (\ref{FineTuning}) gives,

\begin{align}
\mathcal{O}(10^{4}) & = \mathcal{O}(10^{6}) - \mathcal{O}(10^{6})  \nonumber 
\end{align}

\noindent while at the upper end of this range, $\Lambda_{SM} = 10^{19}$ GeV, it gives

\begin{align}
\mathcal{O}(10^{4}) &= \mathcal{O}(10^{38}) - \mathcal{O}(10^{38}) .  \nonumber
\end{align}

\noindent That is, the ``best-case" scenario, where $\Lambda_{SM} = 1 \times 10^{3} \text{ GeV}$, requires that $m_{0}^{2}$ and $\delta m^{2}$ cancel to one part in $10^{2}$ to $10^{3}$. The ``worst-case" scenario, where $\Lambda_{SM} = 10^{19} \text{ GeV}$, requires that $m_{0}^{2}$ and $\delta m^{2}$ cancel to one part in $10^{34}$. For many physicists, cancellation to one part in $10^{2}$ - $10^{3}$ already begins to be problematic.

There exist two conflicting attitudes toward the delicate cancellation between $m_{0}^{2}$ and $\delta m^{2}$ required to recover the measured value of $m_{p}^{2}$ for large values of $\Lambda_{SM}$:

\begin{itemize}
\item This cancellation requires an ``unlikely" ``conspiracy" between the bare Higgs mass and the bare parameters that enter into the calculation of $\delta m^{2}$. The fine tuning problem can be understood as the need to explain the origin of this cancellation in terms of deeper physical theories beyond the Standard Model. 
\item Neither $m_{0}$ nor $\delta m^{2}$ is directly measurable even in principle and these quantities are therefore not physical. This opens the possibility that the delicate cancellations are mere artifacts of mathematical representation, suggesting that the appearance of a coincidence in urgent need of explanation may therefore be illusory. 
\end{itemize} 

\noindent We review arguments in favor of the first view in more detail in Section \ref{FundParams}, which served to establish naturalness as guide to model building in particle physics. The second view has had substantially fewer proponents in the literature on naturalness, but has been advanced in the work of Wetterich and Bianchi/Rovelli \citep{bianchi2010all}, \citep{wetterich1984fine}, \citep{wetterich2012look}. Within the philosophical literature, Williams has more recently adopted this deflationary view of fine tuning formulations of the Higgs naturalness as grounds for characterizing naturalness in terms of inter-scale autonomy rather than fine tuning \citep{williams2015naturalness}. Our discussion here will aim to show how the tension between these two views hinges on whether one physically interprets high-energy EFT's such as the Standard Model as coming equipped with a unique set of ``fundamental parameters."

\section{Motivating Naturalness: Wilsonian Effective Field Theories (EFTs) and ``Fundamental Parameters" } \label{FundParams}

In this section, we examine the role of  the concept of fundamental parameters in motivating the introduction of the naturalness principle. This notion reflects a particular understanding of the Wilsonian approach to renormalization, which takes a Wilsonian EFT to be mathematically defined with a cutoff given by its physical cutoff, and with a specific set of values for its bare parameters, which are understood to be the unique, physically correct parameters of the theory. Below, we will discuss several quotations that reveal the essential aspects of this approach to defining Wilsonian EFTs. In Section \ref{Auxiliary}, we will sketch one alternative possible interpretation of the formalism of Wilsonian renormalization, in which an EFT is not defined uniquely by any single set of values for its bare parameters, or with a single value of the cutoff parameter $\Lambda$, but instead by an entire equivalence class of parametrizations associated with different points along the theory's Wilsonian RG trajectory and different values of $\Lambda$.

\subsection{Naturalness and Fundamental Parameters} 

In his seminal 1979 article on naturalness, Susskind explicitly introduces the notion of  ``fundamental parameters." He writes,

\begin{quote}
The need for fundamental scalar fields in the theory of weak and electromagnetic forces is a serious flaw. Aside from the subjective aesthetic argument, there exists a real difficulty connected with the quadratic mass divergences which always accompany scalar fields. These divergences violate a concept of naturalness which requires \textit{the observable properties of a theory to be stable against minute variations of the fundamental parameters ...} [emphasis ours]

The basic underlying framework of discussion of naturalness assumes the existence of a fundamental length scale $\kappa^{-1}$, which serves as a real cutoff. Many authors have speculated that $\kappa$ should be of order $10^{19}$ GeV corresponding to the Planck gravitational length. \textit{The basic parameters of such a theory are some set of dimensionless bare couplings $g_0$ and masses ... $\mu_0 = \frac{m_0}{\kappa}$. The principle of naturalness requires the physical properties of the output at low energy to be stable against very small variations of $g_0$ and $\mu_0$.} [emphasis ours] \citep{susskind1979dynamics}.
\end{quote}

\noindent In identifying a single set of values for the bare parameters as fundamental, Susskind appears to regard these parameters as serving to define a Wilsonian effective field theory, and to regard all other parametrizations of the theory's amplitudes as derived from this underlying parametrization, and as less fundamental in this sense. Here, the physical cutoff $\kappa$ of the Standard Model is understood to be an essential part of the mathematical specification of the theory. 

Schwartz's recent popular textbook on quantum field theory interprets Susskind as suggesting that the fundamental bare parameters of the Standard Model are physical, and states that ``much of our intuition for fine-tuning and naturalness comes from condensed matter physics." Adopting a Wilsonian formulation of the Standard Model, he writes,

\begin{quote}
Suppose the theory were \textit{finite}, for example if it were UV completed into string theory, or more simply if it were the effective description of some condensed matter system (in which case $\Lambda$ might represent some parameter of the microscopic description, such as the inverse atomic spacing). Then the bare mass $m$ and cutoff $\Lambda$ would be physical. In this situation, we could take the $\Lambda^{2}$ divergence ... literally. ... If the scalar were the Higgs whose pole mass is $m_{p} \approx 125$ GeV, and $\Lambda$ were of the order of the Planck scale, $\Lambda \sim M_{pl} \sim 10^{19}$ GeV, we would need $m^{2} = (1 + 10^{-34}) \Lambda^{2}$. This is called \textbf{fine-tuning}. Fine-tuning is a sensitivity of physical observables (the pole mass) to variation of parameters in the theory \citep{schwartz2014quantum}, Ch. 22.  
\end{quote}

\noindent According to Schwartz's reading of Susskind, bare parameters at the Standard Model's physical cutoff are physical in a sense analogous to the sense in which the parameters of certain lattice models of condensed matter systems, which characterize the interactions between adjacent atoms or molecules in the lattice, are physical. Moreover, these bare parameters are taken to be fundamental in the sense that the theory's predictions follow from them, just as macroscopic behavior of a condensed matter system follows from its microscopic dynamics. However, it is important to emphasize here that Susskind himself makes no reference to analogies with condensed matter theory in his original article introducing the Higgs naturalness principle. It is possible that Schwartz's reading of Susskind, in its emphasis on the role of analogies with condensed matter theory in motivating naturalness, may be injecting an element that Susskind himself did not intend. More careful analysis is needed to fully assess the historical importance of analogies to condensed matter theory in motivating the naturalness principle.

 't Hooft adopts a concept of fine tuning that explicitly emphasizes the ``unlikeliness" of the cancellation necessary to recover observable values in the presence of a fundamental scalar field, which brings with it the notorious quadratic divergences. He likens the relationship between high- and low-energy parameters in a quantum field theory to the relationship between macroscopic and microscopic descriptions of a liquid or solid: 

\begin{quote}
The concept of causality requires that macroscopic phenomena follow from microscopic equations. Thus the properties of liquids and solids follow from the microscopic properties of molecules and atoms. One may either consider these microscopic properties to have been chosen at random by Nature, or attempt to deduce these from even more fundamental equations at still smaller length and time scales. In either case, \textit{it is unlikely that the microscopic equations contain various free parameters that are carefully adjusted by Nature to give cancelling effects such that the macroscopic systems have some special properties} [emphasis ours]. This is a philosophy which we would like to apply to unified gauge theories: the effective interactions at a large length scale, corresponding to a low energy scale $\mu_1$, should follow from the properties at a much smaller length scale, or higher energy scale $\mu_2$, without the requirement that various different parameters at the energy scale $\mu_2$ match with an accuracy of the order of $\frac{\mu_1}{\mu_2}$. That would be unnatural. \citep{hooft1980naturalness}
\end{quote}

\noindent The naturalness requirement as 't Hooft formulates it is motivated by the intuition that the fundamental, microscopic, high-energy parameters do not ``conspire" to give particular  macroscopic, low-energy results. Here, it is the high-energy parameters of a given EFT that are fundamental, in the sense that the values of low-energy parameters follow from them, but not the reverse.

The understanding of fine tuning as sensitivity to variations of fundamental parameters articulated by Susskind is closely tied to the understanding of fine tuning as relying on an ``unlikely" choice of these parameters. 
\footnote{Fine tuning is also associated with sensitivity of observables to fundamental parameters in the work of Barbieri and Giudice, in which they propose quantitative measures of fine tuning that quantify the rate of change of observables with respect to these parameters \citep{barbieri1988upper}. 
}
This can be seen by expressing the Higgs pole mass $m_{p}^{2}$ in terms of the dimensionless bare Higgs mass and top quark Yukawa coupling $y_{t}$ (which is already dimensionless): $m^{2}_{p} \approx \Lambda_{SM}^{2}(\tilde{m}_{0}^{2} - \frac{y_{t}^{2}}{8 \pi^{2}} )$. Taking $\Lambda_{SM} = 10^{19}$ GeV and $m_{p} = 10^{2}$ GeV, we see that a small change in either $\tilde{m}_{0}^{2}$ or $y_{t}^{2}$ of order, say, $10^{-9}$ leads the value of the physical, pole mass $m_{p}^{2}$ to jump by a factor of $10^{27}$. The notion that an ``unlikely" delicate cancellation between $\tilde{m}_{0}^{2}$ and  $\frac{y_{t}^{2}}{8 \pi^{2}}$ is needed to recover the measured value of $m_{p}^{2}$ rests on the fact that $\frac{m_{p}^{2}}{\Lambda_{SM}^{2}}$ is many orders of magnitude smaller than either $\tilde{m}_{0}^{2}$ or $\frac{y_{t}^{2}}{8 \pi^{2}}$; this, in turn, entails that slight changes in either of these dimensionless bare parameters will increase $m_{p}^{2}$ by many orders of magnitude. In short: the presence of delicate (and therefore ``unlikely") cancellations in recovering the pole mass implies delicate sensitivity of the pole mass to slight changes in the values of dimensionless bare parameters.

A final important feature of the understanding of effective field theories in terms of fundamental parameters is that an EFT is defined fundamentally with a cutoff that is equal to its physical cutoff - that is, the UV limit of the energy scales over which the EFT is empirically valid and mathematically defined. Emphasizing this point, Peskin and Schroeder write
\footnote{Thanks to Doreen Fraser for drawing our attention to this quotation. 
}
: 

\begin{quote}
Wilson's analysis takes ... the ... point of view ... that any quantum field theory is defined fundamentally with a cutoff that has some physical significance. In statistical mechanical applications, this momentum scale is the inverse atomic spacing. In QED and other quantum field theories appropriate to elementary particle physics, the cutoff would have to be associated with some fundamental graininess of spacetime, perhaps a result of quantum fluctuations in gravity \citep{peskin1995introduction} Ch. 12, p. 402.
\end{quote}

\noindent On the interpretation of Wilsonian EFT's suggested by Schwartz, it is specifically the fundamental bare parameters defined with respect to this physical cutoff scale that uniquely serve to define the EFT, much as the parameters governing inter-atomic interactions in a condensed matter system define a field-theoretic lattice model of that system, whose physical cutoff scale is equal to the scale associated with the physical inter-atomic lattice spacing..

\subsection{Defining Wilsonian EFTs in Terms of Fundamental Parameters} \label{WilsonFundParams}

Here, we collect and briefly elaborate on several core features of the interpretation of Wilsonian EFT's reflected in the above quotations, which rests on the notion of fundamental parameters. 

\

\noindent \textit{An EFT is mathematically defined with a cutoff equal to its physical cutoff.} In Section \ref{Foundations}, we saw that an EFT can be non-perturbatively defined in the Wilsonian picture with a finite cutoff $\Lambda$ and finite values for the bare parameters $g$. The particular understanding of EFT's suggested by the above quotations (but especially by the quotation from Peskin and Schroeder) states that in defining an EFT mathematically, we should take  $\Lambda = \Lambda_{ph}$, where $\Lambda_{ph}$ is the empirical scale at which the EFT in question ceases to be empirically valid - associated, for example, with the pole mass of a heavy field not included in the EFT Lagrangian, or with some fundamental graininess of spacetime associated with the onset of quantum gravitational effects. For example, QED should by this prescription be defined with a cutoff equal to the mass of the W or Z boson, beyond which the more complete description provided by electroweak theory is needed.  

\

\noindent \textit{An EFT is defined by a single, physical, ``fundamental" bare parametrization.} What does it mean for a set of bare parameters to be ``physical" on this view? Given Schwartz's emphasis on analogies with condensed matter systems, one reasonable interpretation is that this notion of ``physical" requires that it be possible at least in principle to directly, independently measure the values of these parameters, just as one can in principle directly measure the parameters governing inter-atomic or inter-molecular interactions in a condensed matter system, without simply inferring them from the measured values of correlation functions. This physical set of bare parameters $g_{ph}$, together with the cutoff $\Lambda_{ph}$, are then taken as the unique parameters used to define the EFT in the manner described in Section \ref{Foundations}. 

\

\noindent \textit{Wilsonian RG transformations are coarse grainings of the ``fundamental" bare parametrization.} An EFT defined using the physical parameters $g_{ph}$ and $\Lambda_{ph}$ describes degrees of freedom up to the scale $\Lambda_{ph}$, but not above this scale. Wilsonian RG transformations, whereby one integrates out degrees of freedom from $\Lambda_{ph}$ down to some lower scale $\Lambda$ of interest, are then interpreted as coarse graining transformations, which throw away information about physics above the scale $\Lambda$. Thus, along the Wilsonian RG trajectory parametrized by $g(\Lambda)$, there exists one special point $g(\Lambda_{ph}) = g_{ph}$ reflecting the true, physical, ``microscopic" values of the bare parameters.

\

\noindent \textit{Different points along a single Wilsonian RG trajectory parametrize distinct EFTs.} The process of integrating out high-energy modes from $\Lambda_{ph}$ to $\Lambda$ generates a separate, less encompassing EFT, defined only up to the lowered momentum cutoff $\Lambda$, and parametrized by $g(\Lambda)$. Such an EFT \textit{makes no predictions} above the scale $\Lambda$ at which the parametrization $g(\Lambda)$ is defined. The coarse grained parameters $g(\Lambda)$ are related to the parameters $g_{ph}$ of the more fundamental EFT by coarse graining associated with the Wilsonian RG flow to lower momenta. Since the parametrizations $(g(\Lambda), \Lambda)$ associated with different $\Lambda$ are understood to have distinct domains of empirical validity, they are associated with distinct EFTs.

\

More formally, the understanding of Wilsonian EFT's in terms of fundamental parameters takes the physical amplitudes and pole masses predicted by the EFT to be uniquely defined by the fundamental, physical bare parameters $g_{ph} \equiv (g_{1, ph}, g_{2, ph}, ... )$ and physical cutoff $\Lambda_{ph}$: 
  \begin{equation}
  \begin{split}
&\mathcal{Z}[J] = \int^{\Lambda_{ph}} \mathcal{D} \phi \ e^{i \int d^{4} x \left[ \sum_n  g_{n, ph} \mathcal{O}_n  \ + \ J \phi \right]}\,, \\
&G^{(n)}(x_1, ... , x_n; g_{ph}, \Lambda_{ph}) =  \frac{ (-i)^{n}}{\mathcal{Z}[0]} \frac{\delta^n}{\delta J(x_1) \  ... \ \delta J(x_n)}\bigg|_{J=0} \mathcal{Z}[J] \,, \\
& G^{(2)}(x_1, x_ 2; g_{ph}, \Lambda_{ph}) =\\&\qquad=  \int^{\Lambda_{ph}} \frac{d^4 p}{(2\pi)^4} e^{ip(x_1 - x_2)}\left[ \frac{i Z(g_{ph}, \Lambda_{ph})}{p^{2} - m_{p}^{2}(g_{ph}, \Lambda_{ph})} +  \int_{\sim 4 m_{p}^{2}}^{\Lambda_{ph}} d \mu \frac{\rho(\mu)}{p^{2} - \mu^{2}} \right]\,, \\ 
 &S(p_{1}, ... , p_{n}; g_{ph}, \Lambda_{ph}) =
    \tilde{G}^{(n)}(p_{1},... ,p_{n}; g_{ph}, \Lambda_{ph})\ \times
    \\&\hspace*{7em}\times\tilde{G}^{(2), -1}(p_{1}; g_{ph}, \Lambda_{ph})\cdots \tilde{G}^{(2), -1}(p_{n};g_{ph}, \Lambda_{ph}) \,.
 \label{BasicRelationsFund}
  \end{split}
\end{equation} 

\

\noindent All other parametrizations $(g, \Lambda)$, including those related to the fundamental parametrization $(g_{ph}, \Lambda_{ph})$ by Wilsonian RG flows, are derived and less fundamental. Note moreover that on this picture, bare parameters are not a mathematically illicit crutch to be swept under the rug, but are mathematically well-defined and in fact constitute the \textit{defining} parametrization of the EFT. 

A mathematically rigorous, non-perturbative definition of a QFT is often taken to require specification of its Hilbert space and Hamiltonian. If one is willing to relax the requirement of exact Lorentz covariance, there are multiple ways to rigorously define the Hilbert space and Hamiltonian by setting a fixed value for the cutoff and Hamiltonian parameters of the theory. One method is to choose a foliation of spacetime and then place the field degrees of freedom of the theory on a spatial lattice, where the energy scale associated with the inverse lattice spacing $a$ corresponds to the physical cutoff scale of the theory. The Hilbert space of the theory is then the tensor product $\bigotimes_{\vec{x}} \mathcal{H}_{\vec{x}}$ of Hilbert spaces $\mathcal{H}_{\vec{x}}$ associated with different lattice points; the Hamiltonian is a discretized, quantized version of the usual classical field Hamiltonian, with finite values for its coefficients. Alternatively, one can define the Hilbert space as the space of functionals $\Psi[\tilde{\phi}(\vec{k})]$ over classical field configurations containing modes only below the physical cutoff scale; the Hamiltonian of the cutoff theory can be obtained from the continuum Hamiltonian by defining field operators to include only modes below the wave vector associated with the physical cutoff. In both cases, the field operators and parameters defining the Hamiltonian are defined specifically with respect to the physical cutoff scale, as is the Hilbert space. For further discussion of this approach to defining EFTs, see for example \citep{wallaceNaivete} and \citep{ranard2015introduction}.

\vspace{5mm}

The interpretation of Wilsonian QFT's in terms of fundamental parameters has several important virtues by comparison with the pre-Wilsonian understanding of perturbative renormalization, according to which bare parameters were formally infinite and therefore unphysical. 

First, it has the advantage of making the theory mathematically well-defined and avoiding the divergences that arose in older approaches to perturbative renormalization. Every quantity calculated in the theory can in principle be associated with a finite number. Thus, one avoids the conceptual and mathematical obscurity that arises in attempting to deal with the infinite quantities that arise on pre-Wilsonian approaches.  

Second, beyond the question of mathematical well-defined-ness, there may be something deeply intuitively appealing about the view of bare parameters as fundamental parameters, where one has a relatively clear physical picture of the relationship between macroscopic phenomenology and the underlying degrees of freedom described by the theory, as one does in condensed matter applications. By transplanting this intuitive condensed matter picture into the context of elementary particle physics, one similarly gains a sharp (although not necessarily wholly accurate) picture of the underlying degrees of freedom, and of the manner in which the phenomenology of elementary particles emerges from them at a coarse grained level.

\subsection{Higgs Fine Tuning and Fundamental Parameters} 

On the view just described, an EFT is fundamentally specified by a particular set of bare parameters $g_{ph}$ and a particular cutoff $\Lambda_{ph}$, while quantities such as pole masses, scattering amplitudes, and bare parameters $g(\Lambda)$ for $\Lambda < \Lambda_{ph}$  are all understood as deriving from these. The cutoff and bare parameters $g_{ph}$ are thus taken as mutually \textit{independent} parameters in the mathematical definition of an EFT model on this view. Assume that the fundamental parameters are ``chosen by nature" via sampling from a reasonably smooth probability measure over the fundamental bare parameter space. Then the relation $m^{2}_{p} \approx \Lambda_{SM}^{2}(\tilde{m}_{0}^{2} - \frac{y_{t}^{2}}{8 \pi^{2}} )$ suggests that it is overwhelmingly likely, given a large value of the physical cutoff $\Lambda_{SM}$, that a random sampling of $\tilde{m}_{0}^{2}$ and $y_{t}$ will yield a pole mass $m_{p}^{2}$ of order $\Lambda_{SM}^{2}$. That is, ``natural" values for $m_{p}^{2}$ are on the order of $\Lambda_{SM}^{2}$. Only for an extremely atypical and unlikely subset of the bare parameter space of values for $\tilde{m}_{0}$ and $y_{t}$ do there arise the delicate cancellations necessary to recover $m_{p}^{2} << \Lambda_{SM}^{2}$. Recalling the micro/macro analogy of 't Hooft, the sheer unlikeliness of such a choice, as viewed from the perspective of Standard Model effective field theory (SMEFT), suggests a sort of deeper underlying ``conspiracy" between the ``microscopic" bare parameters $\tilde{m}_{0}^{2}$ and $y_{t}$ to yield a particular value for the ``macroscopic" quantity $m_{p}^{2}$.  Put differently, the unlikeliness of the coincidence that must occur for $\tilde{m}_{0}^{2}$ and $y_{t}$ to give small values for the pole mass demands explanation in terms of a deeper, more encompassing BSM theory. The absence of such an explanation within the Standard Model is the feature that is perhaps most often associated with the Higgs naturalness problem. 
\footnote{Building on earlier arguments by Anderson and Casta\~{n}o in \citep{anderson1995measures}, Hossenfelder has recently argued in \citep{hossenfelder2018screams} that the need to assume a smooth probability distribution over the SM parameter space constitutes a weak link in naturalness-based reasoning: what could justify the choice of such a probability distribution, given that the universe appears to sample only one set of values from this space?  While we regard this as an important source of skepticism about fine tuning arguments, we focus here on a separate source of concern as to whether there even exists a unique space of fundamental parameters over which to define this probability distribution. 
}

\section{Deflating Naturalness: Wilsonian EFTs without Fundamental Parameters} \label{Auxiliary}

We now turn to consider the second, deflationary view of Higgs fine tuning described in Section \ref{FineTuningNaturalness}, which regards fine tuning of bare parameters as unproblematic. 

\citep{wetterich1984fine} states that ``fine tuning of bare parameters is not really the relevant problem: we do not need to know the exact formal relation between physical and bare parameters (which furthermore depends on the regularization scheme), and it is not important if some particular expansion method needs fine tuning in the bare parameters or not." Bianchi and Rovelli express a similar point of view in the context of the cosmological constant problem, which is structurally similar to the Higgs naturalness problem, but more severe in that it arises from a delicate cancellation between quartically (rather than quadratically) divergent quantities. They write, ``simple physical arguments indicate that the vacuum energy itself cannot be `real' in the sense of gravitating: if it did, any empty box containing a quantum field would have a huge mass, and we could not move it with a force, since gravitational mass is also inertial mass. On physical grounds, vacuum energy does not gravitate ... A \textit{shift} in the vacuum energy does gravitate." \citep{bianchi2010all} deny that we should ascribe direct physical significance to the value of a bare parameter - in this case, the bare vacuum energy. For this reason, they claim that contrary to the dictates of naturalness, ``there is no great mystery" in the smallness of the cosmological constant. Drawing on Wetterich, \citep{williams2015naturalness} has argued against the formulation of naturalness as a prohibition against fine tuning of bare parameters as follows: ``stuck with an effective theory, one is free to arrange the values of free parameters at high energy - those appearing in the original Lagrangian at the original cutoff scale - however is needed to make accurate predictions for empirically accessible low energy physics, which is all the EFT can reasonably purport to describe anyway. The only concern in doing this is that one isn't fooled into thinking that by being forced to make specific choices for the values of the high-energy parameters, they have thereby learned something meaningful about physics near the cutoff scale."
\footnote{As Williams observes, Wetterich does not deny that naturalness problems exist, but only that, in order to be seen as genuine problems, they should be formulated in terms of physical, renormalized parameters. It is worth noting on this point that renormalized parameters need not always be physical parameters. For example, while the pole mass and renormalized $\overline{MS}$ mass are both renormalized parameters (albeit in different schemes), it is reasonable to question whether the running $\overline{MS}$ mass is genuinely physical in the same sense that the pole mass is: while calculations of the pole mass must yield the same results irrespective of one?s chosen renormalization scheme, and across different effective field theories, the $\overline{MS}$ mass is specific to a particular scheme and its value is not generally preserved across matching between different EFT's. 
}

In characterizing fine tuning of bare parameters as unproblematic, these authors seem implicitly to regard arguments based on the notion of fundamental bare parameters as unconvincing. In particular, Wetterich's claim that fine tuning is attached to a particular expansion method suggests that it is an artifact of arbitrary, unphysical conventions chosen for the purpose of facilitating calculation, akin to a choice of coordinate axes or gauge. In this section, we seek to clarify one possible set of foundations for this view. Specifically, we sketch one way of formulating and interpreting Wilsonian effective field theories that does not require specification of single preferred set of fundamental parameters, and that clarifies more precisely the sense in which fine tuning is merely an artifact of mathematical convention. 

In this way of understanding Wilsonian EFTs, an EFT is specified by an entire Wilsonian RG trajectory rather than by any single point on such a trajectory, and points along the RG trajectory are understood as physically equivalent parametrizations of the same set of physical quantities. One major motivation for such an approach is the fact that correlation functions, from which the observables measured in particle accelerators are calculated, are exactly invariant under re-parametrization by Wilsonian RG flows. This entails that many different parametrizations $(g(\Lambda), \Lambda)$ associated with different values of $\Lambda$ generate exactly the same physical predictions within this context, and that the calculational procedure for generating these predictions therefore does not require that any particular set of parameters be singled out as fundamental, or as the physically correct parametrization of the theory. This in turn suggests that the notion of a single, physically preferred, fundamental bare parametrization may constitute an idle posit in the generation of an EFT's successful empirical predictions, and that it may be possible to go without this assumption in the mathematical formulation of Wilsonian EFTs. Specification of a QFT's correlation functions is often said to encode the full dynamics of the QFT;  ``textbook" QFT, including derivations of the Feynman graph expansion, relies essentially on the calculation of correlation functions. To the extent that all physically relevant dynamical information and structural information about an EFT's state space are encoded in that EFT's RG-invariant correlation functions, it may be possible - by analogy with Wightman's strategy in the context of axiomatic QFT - to build the Hilbert space and Hamiltonian using these correlation functions, and thereby avoid singling out any particular parametrization as fundamental in the mathematical formulation of the EFT.

We now sketch some important features of this way of understanding Wilsonian effective field theory, and then consider several possible objections. 

\

\subsection{Wilsonian EFT's without Fundamental Parameters} \label{NoFundParams}

In this subsection, we describe several core features of this approach to interpreting the Wilsonian RG formalism without singling out any particular set of bare parameter values as fundamental. 

\

\noindent \textit{Physical quantities as Wilsonian RG invariants:} The only quantities calculated within a given EFT that may represent physical quantities - as opposed to quantities that depend on an arbitrary mathematical convention, such as choice of gauge or renormalization scheme - are invariant under re-parametrization by Wilsonian RG transformations. For example, the pole mass and S-matrix elements are admitted as candidates for physicality since they possess this invariance:

\begin{align*}
& \Lambda \frac{d}{d \Lambda} S(p_{1}, ... , p_{N}; g(\Lambda), \Lambda) = 0 \\
& \Lambda \frac{d}{d \Lambda} m^{2}_{p}(g(\Lambda), \Lambda) = 0
\end{align*}

\noindent The invariance of these quantities follows from the Wilsonian RG invariance of the n-point Green's functions:

\begin{align} \label{InvariantGreenFcn}
& \Lambda \frac{d}{d \Lambda} \tilde{G}(p_{1}, ... , p_{n}; g(\Lambda), \Lambda) = 0
\end{align}

\noindent which in turn follows from the invariance of the partition function: $\Lambda \frac{d}{d \Lambda} \mathcal{Z}[J; g(\Lambda), \Lambda] = 0$. 
\footnote{Note that the Callan-Symanzik equation follows directly from (\ref{InvariantGreenFcn}) via application of the Chain Rule. 
}
However, we should note that since S-matrix elements contain an arbitrary global phase, it is rather mod-square S-matrix elements that one might choose to regard as physical. Thus, in the view of Wilsonian EFT's sketched here, Wilsonian RG invariance constitutes a necessary but not sufficient condition for physicality. By contrast, on the ``fundamental parameters" interpretation of Wilsonian EFT models, a single set of values for the cutoff and bare parameters, which are manifestly not invariant under Wilsonian RG flow, are also regarded as physical - i.e., in principle, one could measure them directly and independently, rather than inferring their values from measured values of pole masses and cross sections.

\

\noindent \textit{Wilsonian RG transformations as invertible re-parametrizations:} In the absence of fundamental parameters, Wilsonian RG transformations are regarded as \textit{re-parametrizations} that transform between physically equivalent, finite representations of the same physics, rather than as coarse grainings of a single ``microscopic," high-energy description, associated with the physical-cutoff bare parametrization $g(\Lambda_{ph}) = g_{ph}$. A change of scale $\Lambda$ and parameters $g(\Lambda)$ associated with the Wilsonian RG flow does not signal passage to a more or less fundamental EFT, but merely passage to a different finite parametrization of the \textit{same} EFT. Thus, within a single EFT, one may represent S-matrix elements and pole masses using a low value $\Lambda_{l}$ of the unphysical reference scale $\Lambda$, or a high value $\Lambda_{h}$:

\begin{align*}
S(p_{1}, ... , p_{n}; g(\Lambda_{l}), \Lambda_{l}) &= S(p_{1}, ... , p_{n}; g(\Lambda_{h}), \Lambda_{h})  \equiv  S(p_{1}, ... , p_{n})  \\
m^{2}_{p}(g(\Lambda_{l}), \Lambda_{l}) &= m^{2}_{p}( g(\Lambda_{h}), \Lambda_{h}) \equiv m^{2}_{p}.
\end{align*}

\noindent Thus, it is not the case that bare parameterizations $g(\Lambda_{h})$ referenced to a high cutoff parameter scale $\Lambda_{h}$ are more fundamental than bare parametrizations $g(\Lambda_{l})$ referenced to a low cutoff parameter scale $\Lambda_{l}$, if greater fundamentality is understood to require that the more fundamental description strictly contain the range of phenomena described by the less fundamental description, and describe these phenomena in greater accuracy and/or detail. Contrary to the suppositions of the ``fundamental parameters" view, the low-scale parametrization $(g(\Lambda_{l}), \Lambda_{l})$ captures precisely the same set of physical phenomena as the high-scale parametrization $(g(\Lambda_{h}), \Lambda_{h})$, and is no less physically encompassing. This feature is explained further in subsections \ref{CoarseGraining} and \ref{LowScale}.

\

\noindent \textit{An EFT is specified by an entire Wilsonian RG trajectory, not any single point on such a trajectory:}  On the understanding of Wilsonian RG transformations as mere changes of parametrization within a single EFT, it is more appropriate to associate a given EFT not with any single parametrization, but with the entire continuum of physically equivalent, finite parametrizations that lie along the EFT's Wilsonian RG trajectory. Thus, the EFT is not uniquely associated with any single point along a Wilsonian RG trajectory, but with the entire trajectory itself, which represents a one-parameter equivalence class of mathematically well-defined parametrizations, all of which yield exactly the same set of physical amplitudes. While any single point along a Wilsonian RG trajectory serves to define the whole trajectory via the Wilsonian RG equations, it is unneccessary for the purpose of generating the EFT's successful empirical predictions to single out any particular such point as providing the unique physically correct ``microscopic" parametrization of the theory. Thus, the approach to EFT's without fundamental parameters retains the advantage that the EFT as defined in this manner is finite (since it is defined by an equivalence class of finite parametrizations), but relinquishes the assumption that any single finite parametrization is physically preferred or fundamental. 

What then is the physical significance of the Wilsonian RG flow, given that it merely represents a change of parametrization? In part, the answer lies in the implications of this flow for the variation of S-matrix elements as the physical, external momenta $p_{i}$ are scaled uniformly by some real number $s$:

\begin{align}
S(s p_{1}, ... , s p_{N}; g(\Lambda), \Lambda) =  \left[s^{2-d} \frac{Z(\Lambda)}{Z(s \Lambda)} \right]^{-N/2} S(p_{1}, ... , p_{N}; g(s\Lambda), s\Lambda),
\end{align}

\noindent where $Z(\Lambda) \equiv g_{1}(\Lambda)$. Thus, given the Wilsonian RG trajectory $g(\Lambda)$, the predictions of the theory for $S(p_{1}, ... , p_{N})$ are defined up to scales for which the RG trajectory is mathematically defined (e.g., up to the Landau pole, if one exists). Note that, in cases where the Wilsonian RG trajectories of two theories agree approximately at small $\Lambda$, but diverge for large $\Lambda$, the predictions for $S(p_{1}, ... , p_{N})$ should be approximately equal for small values of the physical scales $p_{i}$ but differ substantially for large $p_{i}$.

\

\noindent \textit{The physical cutoff of an EFT does not occur in the mathematical definition of that EFT:} It is important to note that in defining an EFT by a one-parameter equivalence class of finite parametrizations $(g(\Lambda), \Lambda)$, the physical cutoff $\Lambda_{ph}$ of an EFT appears nowhere in the mathematical definition of that EFT. Without assuming the existence of a preferred set of fundamental parameters, one still obtains the same predictions for physical quantities for any parametrization $(g(\Lambda), \Lambda)$ for which the RG trajectory is defined. Certainly, one is free as a matter of arbitrary convention to set $\Lambda = \Lambda_{ph}$, but nothing is gained in terms of scope or accuracy of the EFT by doing so. The scale $\Lambda_{ph}$ is purely a reflection of the fit between the EFT model and the world, not part of the definition of the EFT; $\Lambda_{ph}$ must be determined empirically. While it is an oft-repeated claim that EFTs predict their own breakdown in the UV, this is true only in the attenuated sense that the Wilsonian RG trajectory may cease to be mathematically defined at some upper limit  $\Lambda^{*}$, as in cases where the theory possess a Landau pole. Beyond the relatively weak requirement that $\Lambda_{ph} < \Lambda_{*}$, nothing about the intrinsic mathematical definition of the EFT implies a specific value for $\Lambda_{ph}$. Indeed, before one can establish $\Lambda_{ph}$ as the physical cutoff of an EFT, one must be able to generate predictions from that EFT that can be compared against experiment. This suggests that the mathematical procedure used to generate the predictions can be defined without reference to $\Lambda_{ph}$. The view of EFTs as specified by an entire RG trajectory rather than by any particular point along that trajectory offers one way of understanding EFTs along these lines. On this view, the physical cutoff $\Lambda_{ph}$ of an EFT is no more intrinsic to the mathematical definition of that EFT than the speed of light $c$ is intrinsic to the mathematical definition of models in Newtonian mechanics.

\vspace{10mm}

Perhaps the strongest motivation for adopting this view of EFTs is that it sheds what appears to be an idle assumption of the view based on fundamental parameters - namely, the very notion that there exists a single, physically preferred bare parametrization. As we have seen, the assumption that there exists a single physically preferred set of values for the bare parameters is not necessary to compute the values of observables like cross sections and pole masses from the Wilsonian path integral expression for correlation functions, since Wilsonian RG invariance of correlation functions entails that many values for these parameters generate exactly the same predictions.

A second argument for shedding the notion of fundamental parameters is that even in cases where it is currently possible in practice to experimentally probe physics at or above the scale of an EFT's physical cutoff, one does not gain the ability to directly measure the theory's bare parameters independently of tuning their values to measured values of correlation functions or observables constructed from correlation functions. 
Recall that we understand bare parameters here as the coefficients $g(\Lambda)$ of the Lagrangian density in the $\Lambda$-cutoff path integral exponent; 
\footnote{Note that it is these parameters that occur in the fine tuning relation \ref{FineTuning}, as can be seen by carrying out a perturbative calculation of the two-point correlation function, which serves to define the pole mass, in the path integral picture. 
}
as such, they are not observables in themselves, since their numerical values are mediated by an arbitrary choice of $\Lambda$ along the RG trajectory. Since many different values of $g(\Lambda)$ for many different values of $\Lambda$ along the same RG trajectory give exactly the same values for all observables predicted by the EFT, measurements of these observables cannot even in principle determine the values of the parameters $g$ uniquely without first specifying an arbitrary value for the scale $\Lambda$. Where the calculation of these observables is concerned, the values of $g(\Lambda)$ are imbued with an element of mathematical conventionality associated with the arbitrariness of $\Lambda$; this is similar to the sense in which the numerical coordinate values of an object are imbued with the conventionality of a choice of origin. Unlike the microscopic parameters of a solid state system, which can be determined either by direct probing of inter-atomic interactions or by fitting correlation functions derived from these parameters, the bare parameters of a particle physics EFT can only be determined by fitting to quantities built from correlation functions, such as cross sections. Thus, there is no independent empirical basis for identifying any particular point along the EFT's RG trajectory as the physically correct, fundamental parametrization of the theory. 
	
	To illustrate more explicitly why one cannot directly measure the bare parameters of an EFT even when one is able to experimentally probe physics near its physical cutoff, imagine EFT1 describing a light field $\phi$ with pole mass $m_{\phi}$, and another more encompassing EFT2 that describes $\phi$ as well as a much heavier field $\psi$ with pole mass $M_{\psi} \gg m_{\phi}$. The heavy particle pole mass $M_{\psi} $ functions as a physical cutoff for EFT1 in that it describes the energy scales at which the predictions of EFT1 cease to be empirically valid (since, for example, EFT1 does not describe production of a $\psi$ particle). 
	The observables of EFT1 include $m_{\phi}$ and $\phi$ field scattering cross sections. The observables of EFT2 include all of the observables of EFT1, as well as the $\psi$ pole mass $M_{\psi}$, $\psi$ field scattering cross sections, and $\phi \psi$ scattering cross sections. The observables of EFT1 are functions $\mathcal{O}_{\phi}(g_{\phi}(\Lambda), \Lambda)$, calculated by differentiation of EFT1's path integral $\mathcal{Z}_{1}[J_{\phi}; g_{\phi}(\Lambda), \Lambda]$ with respect to $J_{\phi}$. Within EFT2, these very same quantities are calculated as functions $\mathcal{O}_{\phi}(g'_{\phi}(\Lambda), g'_{\psi}(\Lambda), g'_{\phi \psi}(\Lambda), \Lambda)$ of the EFT2 parameters by differentiation of the EFT2 path integral $\mathcal{Z}_{2}[J_{\phi}, J_{\psi} ; g'_{\phi}(\Lambda), g'_{\psi}(\Lambda), g'_{\phi \psi}(\Lambda), \Lambda]$ with respect to $J_{\phi}$. For the observables $\mathcal{O}$ of either EFT1 and EFT2, $$\Lambda \frac{d}{d \Lambda} \mathcal{O}(g(\Lambda), \Lambda) = 0\,,$$ so that any bare parametrization along the Wilsonian RG trajectory of either EFT constitutes a viable parametrization of the EFT's observables. 

It is not the case that some particular point along the Wilsonian RG trajectory reveals itself as the true physical parametrization of the EFT when we begin to probe energies around and beyond EFT1's physical cutoff $M_{\psi}$. Rather, what happens at this energy scale is that in addition to the observables $\mathcal{O}_{\phi}$, we begin to probe new observables $\mathcal{O}_{\psi}$ and $\mathcal{O}_{\phi \psi}$, for which EFT1 is inadequate and the more encompassing EFT2 is needed. To describe this larger set of observables, a new set of parameters $(g'_{\phi}(\Lambda), g'_{\psi}(\Lambda), g'_{\phi \psi}(\Lambda))$ is needed, where the EFT2 coefficients $g'_{\phi}(\Lambda)$ of pure $\phi$ field operators include a bare mass parameter $m'_{\phi}(\Lambda)$. The $\phi$ field bare mass $m'_{\phi}(\Lambda)$ in EFT2 will in general not be equal to the $\phi$ field bare mass $m_{\phi}(\Lambda)$ in EFT1, although the $\phi$ field pole mass $m_{\phi}$ will be equal between the two theories. It is precisely in order to keep the values of observables like the pole mass $m_{\phi}$ fixed that it is necessary to perform ``matching corrections" that relate the bare masses in EFT1 and EFT2 $m'_{\phi}(\Lambda) = m_{\phi}(\Lambda) + \delta m_{\phi}(\Lambda)$. Thus, the value of the $\phi$ field bare mass depends both on the model being used - EFT1 or EFT2 - and on the arbitrary choice of $\Lambda$ within that model, by contrast with the pole mass $m_{\phi}$, which depends on neither. The dependence of the bare mass on such arbitrary modeling choices provides one reason to take the bare mass less physically seriously than the pole mass. 

On the other hand, it is true that once one has ascertained the parameter values for EFT2 by fitting to the observables $\mathcal{O}_{\phi}$, $\mathcal{O}_{\psi}$, and $\mathcal{O}_{\phi \psi}$, one can then derive the values of the parameters of EFT1 on the assumption that EFT1 yields the same values for the observables $\mathcal{O}_{\phi}$ as those predicted by EFT2. Thus, it may appear that there are indeed two independent ways of measuring the $g_{\phi}(\Lambda)$: one may either fit the parameters $g_{\phi}(\Lambda)$ of EFT1 directly to the measured values of the observables $\mathcal{O}_{\phi}$, or one may fit the parameters of EFT2 to the set  $\mathcal{O}_{\phi}$, $\mathcal{O}_{\psi}$, and $\mathcal{O}_{\phi \psi}$, and then derive the EFT1 parameters via the matching requirement that they approximately recover the predictions of EFT2 for $\mathcal{O}_{\phi}$. However, these two ways of ascertaining $g_{\phi}(\Lambda)$ are not independent, since both ultimately involve fitting the same set of parameters $g_{\phi}(\Lambda)$ to the same set of measurements of $\mathcal{O}_{\phi}$. The latter method is simply a more roundabout way of fitting $g_{\phi}(\Lambda)$ to measurements of $\mathcal{O}_{\phi}$. 

We can if we like set $\Lambda = M_{\psi}$ in EFT1, which may seem a natural move since both $M_{\psi}$ and $\Lambda$ are identified as the ``cutoff" of EFT1. But there is no particular advantage to doing so; we do not arrive at a more encompassing theory with a more encompassing set of observables, or at a more accurate prediction of these observables, by virtue of this choice of $\Lambda$ rather than smaller values. Although both are called the ``cutoff" of EFT1, the parameter $\Lambda$ in EFT1 and the pole mass $M_{\psi}$ in EFT2 are entirely distinct types of quantity and should not be conflated: the second is a concrete physical quantity, invariant under Wilsonian RG transformations of the bare parameters in EFT2; the first is an unphysical parameter that sets the arbitrary scale relative to which the parametrization $g_{\phi}(\Lambda)$ of EFT1 is defined. Particularly in cases where we can probe physics at and above $M_{\psi}$, the bare parameters $g_{\phi}(\Lambda = M_{\psi})$ along the defining RG trajectory of EFT1 do not represent new physics beyond EFT1, even though they are sometimes interpreted this way. The $g_{\phi}(M_{\psi})$ are just another parametrization of EFT1 observables, among a continuous infinity of physically equivalent parametrizations $g_{\phi}(\Lambda)$. They give us no more (or less) information about physics beyond the realm of EFT1 - for example, about the observables $\mathcal{O}_{\phi \psi}$ and $\mathcal{O}_{\psi}$ of EFT2 - than do the EFT1 bare parameters $g_{\phi}(\Lambda)$ at any other value of $\Lambda$ (even $\Lambda =0$).

It is worth noting that the case of matching between EFT1 and EFT2  just described does not reflect the most general type of relationship between two EFTs, since there are cases where the fields described by the low-energy EFT1 do not occur in the more encompassing EFT2's Lagrangian, but are instantiated by  combinations of the fields and degrees of freedom of EFT2. This occurs for example in the relationship between chiral perturbation theory and QCD. However, the grounds for our argument that the bare parameters cannot be measured even when it is possible to probe physics near the scale of the low-energy EFT's physical cutoff is not altered in this case. In this case, as in the case just discussed, the Lagrangian coefficients that appear in the finite-cutoff path integral expression for each EFT depend on an arbitrary choice of $\Lambda$, and are not themselves observables.
	
However, forgetting for the moment about bare parameters, this type of case does present the possibility of genuinely problematic fine tuning - for example, in the case where the Higgs scalar turns out to be a composite particle made up of much heavier particles described by some BSM theory. In this case, a light Higgs pole mass requires delicate cancellation between the much larger pole masses of its constituents and their binding energy. Unlike the cancellation between the bare Higgs mass and quantum corrections, which we argue is an artifact of a certain conventionally chosen parametrization scheme, such a cancellation would involve quantities that are genuinely physical. However, to date there is no evidence to suggest that the Higgs scalar is a composite of heavy particles described by some deeper BSM theory.

A third point in favor of shedding the notion of fundamental parameters applies would hold in scenarios for physics beyond the Standard Model where the Higgs pole mass enters as a basic rather than derived input to the fundamental theory - let us call it $T_{new}$. In such cases, it cannot be true that the bare parameters $g(\Lambda_{phys})$ of the Standard Model at its physical cutoff are physically more fundamental than the Higgs pole mass. It makes little sense to countenance $m_{p}^2$ as fundamental in $T_{new}$ but as non-fundamental in SMEFT, given that SMEFT is only a low-energy approximation to $T_{new}$. In this case, it is clear that the bare Higgs mass $m^{2}(\Lambda_{SM})$ is not fundamental or
physical, but is artificially tuned to whatever value it must possess to recover the measured values of genuinely physical quantities such as the Higgs pole mass. 

A fourth  argument is that since the view of EFTs as defined by a set of fundamental parameters takes the state space and dynamics of an EFT to be truncated in the UV by $\Lambda_{ph}$, the EFT makes no predictions about phenomena above the scale of $\Lambda_{phys}$. The state space simply does not possess the states to describe scattering processes at physical scales (e.g. center-of-mass energy, momentum transfer) above $\Lambda_{phys}$. For this reason, the view of EFTs based on fundamental parameters seems inconsistent with the conventional notion that QED is also capable of making predictions (albeit not empirically correct ones) about phenomena above the scale of its physical cutoff. The physical cutoff is determined by comparison of the EFT's predictions with experiment, suggesting that the EFT is capable of generating predictions independently of our knowledge of the physical cutoff, and that it may therefore be possible to define the EFT in a way that does not make essential reference to the physical cutoff. The view of EFTs as defined by an entire Wilsonian RG trajectory $g(\Lambda)$ for arbitrary $\Lambda$, rather than by the single point $g(\Lambda_{ph})$, fits more naturally with the notion that it should be possible to define the theory in a way that does not reference its physical cutoff, since an EFT in this formulation makes predictions for scattering amplitudes at all scales for which the RG trajectory is defined, and not just up to the physical cutoff scale.

\vspace{5mm}

We now consider several possible concerns about the viability of the interpretation of Wilsonian EFTs just sketched.

\subsection{Objection 1: Aren't Wilsonian RG Transformations Coarse Graining Transformations?} \label{CoarseGraining}

The Wilsonian RG flow from high to low scales is often regarded as a coarse graining transformation. However, this point requires some care, since the Wilsonian RG equations,

\begin{align} \label{WilsonRGNoFund}
\Lambda \frac{d}{d \Lambda} g_{i}(\Lambda) = \beta_{i}(g(\Lambda), \Lambda)
\end{align}  

\noindent are first-order in $\Lambda$ and therefore invertible. As Morris has emphasized, there is no ``loss of information" in the Wilsonian RG flow from a parametrization $g(\Lambda_{h})$ defined at some high scale $\Lambda_{h}$ to the corresponding parametrization $g(\Lambda_{l})$ at a low scale $\Lambda_{l}$ \citep{morris1994exact}, \citep{morris1998elements}. One can run the parameters $g(\Lambda)$ up as well as down in $\Lambda$. 

What then of the notion that the Wilsonian RG explains why theories differing in their description of physics at high-energies all converge to the same renormalizable theory at low energies? Surely information about the details of physics at high energies is lost in transition to describing low-energy phenomena by a renormalizable theory, so that the Wilsonian RG flow can be regarded as coarse graining. Reconciliation can be found in the fact that information is lost only in the approximation where small coefficients of irrelevant operators are ignored, which entails a change to a distinct RG trajectory. Thus, it is not the Wilsonian RG flow in itself that constitutes coarse graining, but this flow combined with the added step of throwing away information about these small coefficients. Prior to performing this second step, the Wilsonian RG flow can be regarded as an invertible re-parametrization of a fixed set of physical amplitudes. 

It is worth noting here that while the process of integrating out infinitesimal momentum shells for a single fixed field or set of fields is invertible in the absence of approximations, the process of integrating out an entire field from the path integral in order to obtain the path integral for an EFT describing some smaller set of fields is \textit{not} generally invertible and so \textit{can} be regarded as a coarse graining.  Starting from a theory of two fields, one light field $\phi$ and one heavy field $\psi$ - where ``light" and ``heavy" are understood in terms of the pole masses of the fields -  one can define an EFT describing the behavior of only the light field at low energies (where particles of the heavy field are not created), by ``integrating out" the heavy field $\psi$ from the path integral of the full theory. The full theory describing both fields $\phi$ and $\psi$ - which may itself be a low-energy approximation to some still more encompassing theory - is more fundamental than the EFT describing only the light field $\phi$ in the sense that the full theory circumscribes the domain of empirical validity of the light field EFT; the light field EFT thus ``reduces to" the full theory. 
\footnote{Within the philosophical literature, the subjects of reduction and emergence as they relate to renormalization and effective field theory have been examined at length in the work of Butterfield, Hartmann, Castellani, Bain, Crowther, Huggett, Williams, Franklin, and others \citep{butterfield2014reduction}, \citep{hartmann2001effective}, \citep{castellani2002reductionism}, \citep{bain2013emergence}, \citep{crowther2015decoupling}, \citep{huggett1995renormalisation}, \citep{williams2017scientific}, \citep{franklin2017whence}. Recent discussion concerning the general methodology of reduction in physics can be found in Butterfield, Crowther, Fletcher, and Rosaler \citep{butterfield2011emergence}, \citep{butterfield2011less}, \citep{butterfield2012emergence}, \citep{butterfield2014reduction}, \citep{fletcher2015similarity}, \citep{crowther2015decoupling}, \citep{crowther2017inter}, \citep{rosaler2015local}, \citep{rosaler2017reduction}, \citep{rosaler2018generalized}, \citep{RosalerThesis}. 
}
Consequently, the parameters  $g_{\phi}(\Lambda), g_{\phi \psi}(\Lambda), g_{\psi}(\Lambda)$ of the full theory (where $g_{\phi \psi}(\Lambda)$ correspond to operators that couple $\phi$ and $\psi$) \textit{do} constitute a more fundamental parametrization of light-field amplitudes than do the parameters $g'_{\phi}(\Lambda)$ of the light field EFT. 
\footnote{Note that the parameters $g'_{\phi}(\Lambda)$ of the light field EFT differ from the light field parameters $g_{\phi}(\Lambda)$ of the full theory since they are subject to so-called ``matching corrections" that reflect the influence of heavy fields that have been integrated out. 
}
Thus, the progression from a more to a less fundamental parametrization occurs not in the flow to smaller values of $\Lambda$, but in the projection of the RG flow of the full theory onto the RG flow of the light field EFT, which in principle can be performed at arbitrary $\Lambda$.

\subsection{Objection 2: Doesn't Larger $\Lambda$ Imply a More Encompassing Theory?} \label{LowScale}

One assumption of the interpretation of Wilsonian EFT's based on fundamental parameters is that a theory parametrized at scale $\Lambda$ only describes phenomena at physical scales below $\Lambda$. Thus, parametrizations $g(\Lambda_{h})$ with respect to a high cutoff $\Lambda_{h}$ describe a distinct, more fundamental theory than do parametrizations $g(\Lambda_{l})$ with respect to a lower cutoff $\Lambda_{l}$. While nothing prevents one from defining EFT's in this manner, such an interpretation is not necessitated by the mathematical formalism of the Wilsonian RG. Other interpretations of this formalism are also possible. In particular, in the interpretation sketched in subsection \ref{NoFundParams}, \textit{any} point $g(\Lambda)$ along the Wilsonian RG trajectory can be used to parametrize the full set of physical amplitudes predicted by that EFT. The unphysical scale parameter $\Lambda$ is independent of the physical scales of the problem (e.g., pole masses, external momenta $p_{i}$), and there is no special need for $\Lambda$ to lie above (or below) these physical scales. 

One can see this by noting that 

\begin{equation*}
\Lambda \frac{d}{d \Lambda} \mathcal{Z}[J; g(\Lambda), \Lambda] = 0,
\end{equation*}

\noindent where $g(\Lambda)$ is a solution to the Wilsonian RG equations (\ref{WilsonRGNoFund}), so that we can write $\mathcal{Z}[J; g(\Lambda), \Lambda] = \mathcal{Z}[J]$. That is, $\mathcal{Z}[J; g(\Lambda), \Lambda]$ is the same function of $J$ irrespective of the value chosen for $\Lambda$. In particular, the functional derivatives of $\mathcal{Z}[J]$ with respect to $J$,

\begin{equation*}
\tilde{G}^{(n)}(p_1, ... , p_n) =  \frac{ (-i)^{n}}{\mathcal{Z}[0]} \frac{\delta^n}{\delta \tilde{J}(p_1) \  ... \ \delta \tilde{J}(p_n)}\bigg|_{J=0} \mathcal{Z}[J],
\end{equation*}

\noindent do not depend on $\Lambda$,

\begin{align*}
 \Lambda \frac{d}{d \Lambda} G^{n}(p_{1}, ... , p_{n}; & g(\Lambda),  \Lambda ) = 0, \\
& \\
\end{align*}

\noindent so that one can write,

\begin{align*}
G^{(n)}(p_{1}, ... , p_{n}; g(\Lambda), \Lambda ) &= G^{(n)}(p_{1}, ... , p_{n}) \\
S(p_{1}, ... , p_{N}; g(\Lambda), \Lambda) &= S(p_{1}, ... , p_{n}) \\
m^{2}_{p}(g(\Lambda), \Lambda) &= m^{2}_{p}. \\
\end{align*}

\noindent As Wilsonian RG invariants, the quantities $G(p_{1}, ... , p_{N})$, $S(p_{1}, ... , p_{N})$, and $m^{2}_{p}$ belong to the set of quantities that may potentially be regarded on this interpretation as ``physical." These equations illustrate a sense of cutoff independence that does not require the existence of a continuum limit, where $\Lambda$ is taken to infinity - rather, the explicit cutoff dependence of quantities such as S-matrix elements and pole masses is removed at \textit{finite} $\Lambda$ by the cutoff dependence of the parameters $g(\Lambda)$. 

The parameter $\Lambda$ can be interpreted as an unphysical reference scale, which sets the boundary between those modes $\tilde{\phi}(k)$ in the path integral that have been explicitly integrated over, with their influence absorbed into the Lagrangian parameters $g(\Lambda)$, and those that have not been explicitly integrated over, so that they appear explicitly in the Lagrangian of the path integral. 
One can understand $\Lambda$'s role in partial analogy to the simple case of bivariate integration:

\begin{align}
F[j] \equiv \int dx \ dy \ f(x,y; j) =  \int dx \ g(x;j)
\end{align}

\noindent where $g(x;j) \equiv \int dy \ f(x,y; j)$. The function $F[j]$ and its derivatives with respect to $j$ are unchanged by whether we write $F[j]$ as the integral over $dx \ dy$ of $f(x,y;j)$ or as the integral over $dx$ of $g(x;j)$. Likewise, the functional $\mathcal{Z}[J]$ and its derivatives with respect to $J$ are unchanged by whether we write $\mathcal{Z}[J]$ as a functional integral over modes below $\Lambda_{1}$ of $e^{i S(g(\Lambda_{1}))}$ or as a functional integral over modes below $\Lambda_{2}$ of $e^{i S(g(\Lambda_{2}))}$ (where $\Lambda_1 \neq \Lambda_2)$. The parameter $\Lambda$, which simply reflects how we choose to write down the expression for the path integral, is thus unrelated to the physical scale set by the external momenta $p_{i}$, and may lie above or below them. For this reason, parametrizations associated with larger $\Lambda$ need not be identified with distinct, more encompassing theories. It is reasonable to regard them instead as alternative parametrizations of a single EFT, since all such parametrizations represent one and the same set of physical amplitudes, derived from the same cutoff-independent function $\mathcal{Z}[J]$. On this interpretation, the parameter $\Lambda$ should therefore not be understood as a cutoff in the particular sense that it truncates the Hilbert space or the set of Fourier modes over which the dynamics of the theory are defined.

\

From the above considerations, it follows that we can use any parameters $g(\Lambda)$ along the Wilsonian RG flow -- not just $g(\Lambda_{phys})$, and not just $g(\Lambda)$ for $\Lambda > p_{i}, m_{p}$, etc. -- to parametrize correlation functions and physical quantities derived from them. More formally, the equations that define an EFT's correlation functions and S-matrix elements in the absence of fundamental parameters take the form,  
\begin{equation}
  \begin{split}
& \mathcal{Z}[J] = \int^{\Lambda} \mathcal{D} \phi \ e^{i \int d^{4} x \left[ \sum_n  g_{n}(\Lambda) \mathcal{O}_n  \ + \ J \phi \right]}\,, \\
& G^{(n)}(x_1, ... , x_n; g(\Lambda), \Lambda) =  \frac{ (-i)^{n}}{\mathcal{Z}[0]} \frac{\delta^n}{\delta J(x_1) \  ... \ \delta J(x_n)}\bigg|_{J=0} \mathcal{Z}[J] \,, \\
& G^{(2)}(x_1, x_ 2; g(\Lambda), \Lambda) =\\&\qquad=  \int^{\Lambda} \frac{d^4 p}{(2\pi)^4} e^{ip(x_1 - x_2)}\left[ \frac{i Z(g(\Lambda), \Lambda)}{p^{2} - m_{p}^{2}(g(\Lambda), \Lambda)} +  \int_{\sim 4 m_{p}^{2}}^{\Lambda} d \mu \frac{\rho(\mu)}{p^{2} - \mu^{2}} \right]\,, \\ 
    & S(p_{1}, ... , p_{n}; g(\Lambda), \Lambda) =
    \tilde{G}^{(n)}(p_{1}, ... ,p_{n}; g(\Lambda), \Lambda)\ \times\\
    &\hspace*{7em}\times\tilde{G}^{(2), -1}(p_{1}; g(\Lambda), \Lambda) \cdots
    \tilde{G}^{(2), -1}(p_{n}; g(\Lambda), \Lambda)\,,
  \end{split}
\end{equation}

\

\noindent where \textit{any} $\Lambda$ for which the Wilsonian RG flow is defined may be used to parametrize the theory. As Morris has noted, one can even take $\Lambda \rightarrow 0$ without changing the values of the correlation functions, or any of the quantities derived from them. By contrast with the formulation given in (\ref{BasicRelationsFund}), no single parametrization is ``fundamental."

\subsection{Objection 3: What are the Hilbert Space and Hamiltonian?} \label{Hilbert}

Perhaps the most salient criticism of the approach that we have sketched is that it is under-formulated. A rigorous mathematical definition of a QFT (including an EFT) is widely understood to require specification of the theory's Hilbert space and unitary time evolution via some Hamiltonian. Such a definition permits application of the theory outside of the restricted context of scattering experiments, as required for example in the context of quantum cosmology (see, e.g., \citep{nelson2017classical}). How are the Hilbert space and Hamiltonian of an EFT specified on the approach just described, in which all points along the RG trajectory constitute physically equivalent parametrizations of observables and physical amplitudes? 

We do not currently have a well-developed answer to this question. Thus, the understanding of Wilsonian EFTs sketched in this section requires further development before it can be regarded as an established alternative to the formulation based on fundamental parameters, where there do exist mathematically well-defined ways to specify the Hilbert space and unitary dynamics of the theory in a general, non-perturbative setting. On the other hand, we know of no result that precludes the possibility of defining a QFT  by its full RG trajectory rather than by any particular point or subset of points along it. Moreover, we think that there exist good motivations for pursuing such a formulation. In particular, as we have emphasized, the actual practice of generating predictions from a path integral Lagrangian - whether this Lagrangian is renormalizable or non-renormalizable - does not require one to specify any single set of parameter values as fundamental since every point along the EFT's RG trajectory generates exactly the same predictions. This suggests that the specification of a single preferred bare parametrization may be superfluous to the theory's empirical success, and motivates the search for a way of formalizing the theory that does not rely on the assumption of such a preferred parametrization. In our view, the absence of any need for fundamental parameters in generating the theory's successful predictions constitutes compelling, but not conclusive, evidence for the possibility of formulating EFTs in manner that treats different parametrizations on more equal footing.

\subsection{Higgs Fine Tuning without Fundamental Parameters} \label{AuxiliaryNaturalness}

On shedding the assumption that a single set of values of the parameters $g$ constitute a physically preferred set of fundamental parameters, the bare parameters at a given scale are no longer mutually independent.  Instead, they are constrained to lie within a one-parameter class of physically equivalent finite parametrizations, each associated with a different value of the unphysical scale parameter $\Lambda$, all of which give the same values for the Higgs pole mass and correlation functions. Noting that the bare Higgs mass $m_{0}^{2}(\Lambda)$ and Yukawa coupling $y_{t}(\Lambda)$ can be interpreted as two of the parameters $g(\Lambda)$ in the cutoff path integral Lagrangian for the Standard Model, the leading contribution to the one-loop expansion of the pole mass at $\Lambda$ gives:

\begin{align} \label{FineTuningAux}
m^{2}_{p} &= m_{0}^{2}(\Lambda) + \delta m^{2} \nonumber(\Lambda) \\
& = m_{0}^{2}(\Lambda) - \frac{y_{t}^{2}(\Lambda)}{8 \pi^{2}} \Lambda^{2}+ \ldots \nonumber \\
& = \Lambda^{2}\left(\tilde{m}_{0}^{2}(\Lambda) - \frac{y_{t}^{2}(\Lambda)}{8 \pi^{2}} \right) + \ldots 
\end{align}

\noindent where $\tilde{m}_{0}^{2} = \frac{m_{0}^{2}}{\Lambda^{2}}$. The scale parameter $\Lambda$ is chosen as a matter of convention. Large choices of $\Lambda$ require more delicate cancellations between $m_{0}^{2}(\Lambda)$  and $\frac{y_{t}^{2}(\Lambda)}{8 \pi^{2}}$, while smaller values require less delicate cancellations. Thus, the delicateness of the cancellation is entirely an artifact of convention as well. In particular, the choice in (\ref{FineTuning}) to set $\Lambda$ to the physical cutoff scale $\Lambda_{SM}$ of the Standard Model reflects an arbitrary (and not particularly convenient) choice of convention that has no bearing on the scope of the Standard Model. The delicate cancellations can be eliminated by re-parametrizing the theory in terms of smaller values of $\Lambda$, which entails no loss of information or scope of the theory. In the absence of fundamental parameters, the claim that the bare parameters $g(\Lambda_{SM})$ at the SM's physical cutoff are more fundamental than those at lower cutoff is severely weakened, since both parametrizations describe exactly the same physics, and each can be inferred from the other. In particular, SMEFT as described by the bare parameterization $(\tilde{m}_{0}^{2}(m_{p}), y_{t}^{2}(m_{p}))$, where $m_{p}$ is the Higgs pole mass, is no less fundamental than SMEFT as described by the bare parametrization $(\tilde{m}_{0}^{2}(\Lambda_{SM}), y_{t}^{2}(\Lambda_{SM}))$,  even though $\Lambda_{SM} >  m_{p}$. 

\begin{figure}
\includegraphics[width=1.0\textwidth]{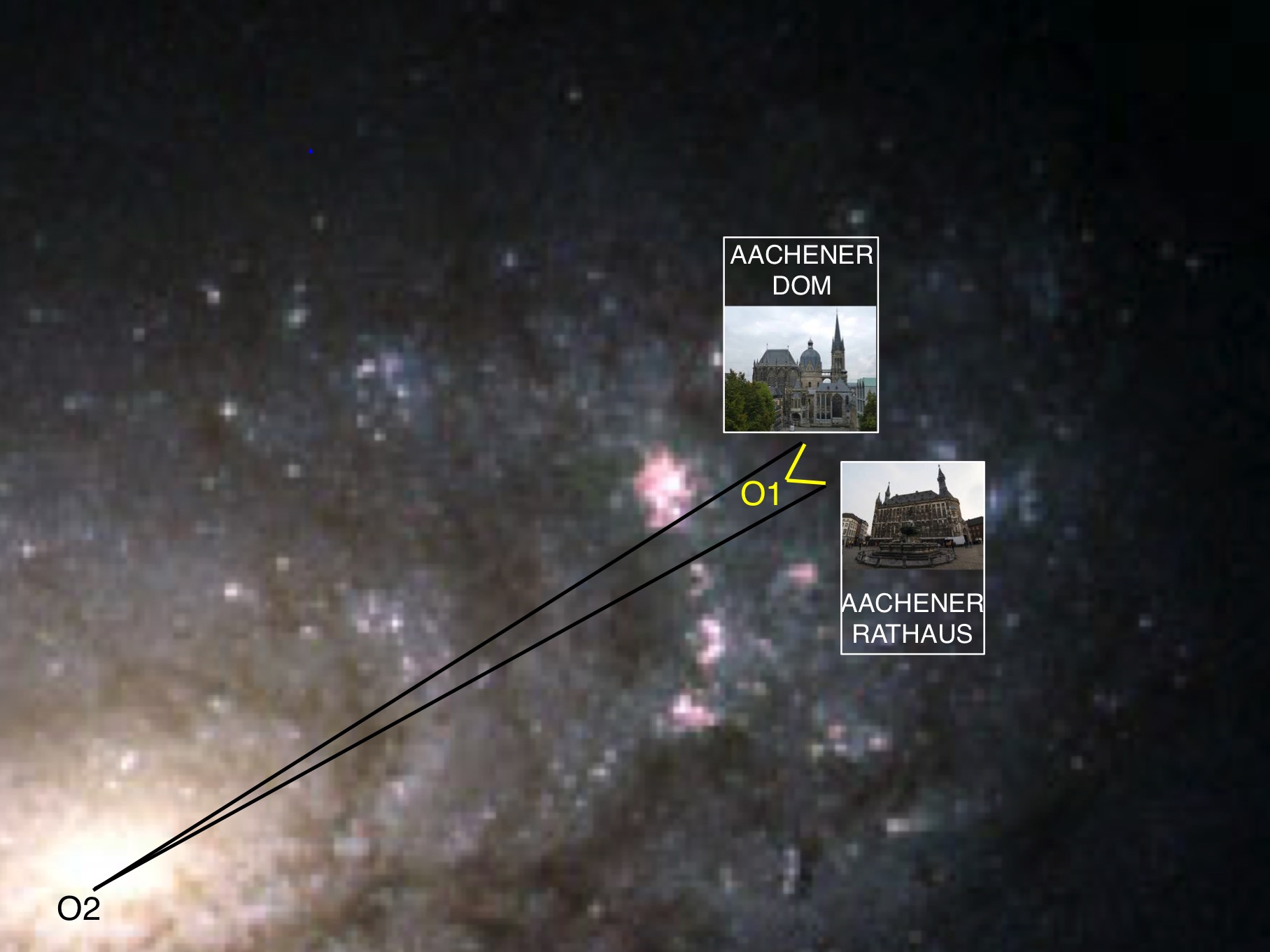}
\caption{With choice of origin at the center of the Milky Way (O2), there is an extremely delicate cancellation - to roughly one part in $10^{18}$ - between vectors indicating the locations of Aachen Dom (Cathedral) and Aachen Rathaus (Town Hall). However, this cancellation can be dramatically reduced by moving the origin close to, say, the midpoint between the two (O1). Likewise, the cancellation between the bare Higgs mass and its quantum corrections is very delicate for large $\Lambda$ (say, the Planck scale), and much smaller for $\Lambda$ on the order of the Higgs pole mass. Without fundamental parameters, the choice of the unphysical reference scale $\Lambda$ is purely conventional and akin to a choice of origin; the delicate cancellation between bare Higgs mass and quantum corrections is then an eliminable, unphysical artifact of convention.} \label{MilkyWay}
\end{figure}

On the view of bare parameters as conventional, regarding the delicate cancellations between $\tilde{m}_{0}^{2}(\Lambda_{SM})$ and $\frac{y_{t}^{2}(\Lambda_{SM})}{8 \pi^{2}}$ as coincidence is akin to finding coincidence in the following delicate cancellations:

\begin{itemize}
\item We wish to determine the mass of an ant by first measuring the mass of the system (ant + X), then measuring the mass of the system (X) alone, and finally subtracting the latter from the former. Taking X=earth, we find that the mass of the system (ant + earth) agrees with the mass of the system (earth) to one part in $10^{30}$.
\item We wish to measure the distance from the Aachen Dom (Cathedral) to the Aachen Rathaus (Town Hall). We do so by measuring the magnitude of the difference between vectors  $\hat{x}_{dom}$ and $\hat{x}_{rat}$ describing the locations of these two buildings, where the vector components are specified relative to a coordinate system whose origin lies at the center of the Milky Way, and whose x-axis lies along the line joining the center of the Milky Way to central Aachen. We find that the x-components of $\hat{x}_{dom}$ and $\hat{x}_{rat}$ agree to one part in $10^{18}$ (see Figure \ref{MilkyWay}). 
\end{itemize}

\noindent In both cases, the delicate agreement between the quantities in question results from an inconvenient, conventional choice of reference point, not a mysterious physical coincidence. In the first example, we might have chosen instead to measure the ant's mass by subtracting the mass of (pencil) from the mass of (ant + pencil), which would have exhibited a much less severe cancellation. The choice of reference system (pencil) or (earth) is arbitrary and conventional, while the quantity of interest - the mass of the ant - is invariant across different choices of reference system. Likewise, in the case of the Higgs, we might express the invariant Higgs pole mass as a difference between bare mass and quantum corrections defined with respect to different conventionally chosen values of $\Lambda$.

\subsection{Scientific Realism and Wilsonian EFT's}

We have argued here that the validity of at least one set of naturalness-based arguments turns critically on questions of physical interpretation of effective field theories - that is, on questions concerning which elements of the mathematical formalism of an EFT may represent real features in the world, and which reflect mere choices of mathematical convention or artifacts of mathematical representation. Such questions are especially salient in the context of efforts to formulate realist interpretations of quantum field theory, which attempt to understand our theories not merely as toolkits for prediction (as operationalist or empiricist approaches tend to do), but as representing - if only in an approximate and domain-restricted way - the structure of the physical world beneath the surface of directly observable phenomena. 
\footnote{The difference between realist and empiricist approaches to the interpretation of scientific theories is often illustrated by reference to debates in the 19th century concerning the existence of atoms - are atoms fictitious theoretical constructs that are merely useful for describing the observable behavior of gases and other physical systems, or do they really exist? The empiricist or instrumentalist view of atoms (which has declined markedly in popularity) adopts the first view, while the realist view adopts the latter. For more detailed analysis of this debate, see for example Psillos' \citep{psillos2011moving}. 
}
Within efforts to give a realist interpretation of quantum field theory, there is the further question of which particular aspects of an EFT's formalism one should be realist \textit{about}, given the observed empirical success of the EFT. 
\footnote{Efforts to interpret effective field theories, and quantum field theories generally, from an empiricist point of view can be found, for example, in Butterfield and Bouatta's \citep{butterfield2015renormalization}, Ruetsche's \citep{ruetsche2011interpreting}, Fraser's \citep{fraser2011take}, and Halvorson and M{\"u}ger's \citep{HalvorsonSynSem}. 
}

Here, we wish to acknowledge one recent line of inquiry into realist interpretations of EFT's in high energy physics, and to briefly orient our own discussion with respect to the point of view considered in this work. Williams, J. Fraser, and Ruetsche have recently explored the notion that the renormalization group may be used as a guide in attempting to uncover which mathematical features of empirically successful EFT's to regard as representing real features in the world, which are likely to be preserved in future, more fundamental theories. Invoking Wimsatt's criterion of robustness for realist commitment, Williams writes, ``What the RG shows is that the `fundamental' short-distance structure ... is largely irrelevant to the physical content of an EFT in the domain where we have any reason to consider it empirically reliable ... An EFT at long distances is `robust' in a way that the the short distance `fundamental' theory is demonstrably not: its entities and structures at that scale are `accessible (detectable, measurable, derivable, definable, producible, or the like) in a variety of independent ways,' and so are candidates for being included in the ontology of that EFT" \citep{williams2017scientific}, \citep{wimsatt2007re}. In a similar vein, Fraser writes, ``in demonstrating that the large scale properties of a QFT model are insensitive to the high energy dynamics, the renormalisation group is also telling us that these properties are essentially independent of the details of future physical theories which describe the dynamics of currently inaccessible high energy degrees of freedom. Thus the renormalisation group gives us a way of identifying properties of our present theories which will be embedded within future theories, in one way or another" \citep{fraser2016quantum}. Thus, both Fraser and Williams see the RG as facilitating the interpretation of EFT's by identifying those low-energy features that are robust against variations in the unknown high-energy physics, and therefore likely to be preserved irrespective of what this new physics turns out to be. 

On the other hand, Ruetsche writes that the arguments offered by Williams and Fraser in support of this view are ``compromised by a certain faux generality afflicting RG analyses." She explains, ``The concern is that even explicit RG results are only as reassuring as the space of theories on which the RG group acts is comprehensive. But that space incorporates assumptions about what kinds of interactions are possible and how to model them. And nature isn't beholden to respect those assumptions" \citep{ruetscheAscent}. In particular, proofs based on the RG that many distinct theories with different values of irrelevant operator coefficients are well approximated at low energies by the same renormalizable theory assume that both the fundamental high-energy theory and approximate low-energy theory reside in the same parameter space describing the same set of fields. On the other hand, Ruetsche suggests, the space of possible high-energy theories may extend even beyond the realm of quantum field theory, and so is much larger than the set of possibilities represented by the parameter space that hosts the Wilsonian RG flow.

The interpretation of Wilsonian EFTs without fundamental parameters, which restricts realist commitment to Wilsonian RG invariants, is motivated by many of the considerations that drive the realism of Williams and Fraser; however, there are several important distinctions that we wish to draw here between the particular realist approach to interpreting QFT's that is advocated by these authors and the brand of realism that informs the view of EFT's sketched in this section. We agree with Williams and Fraser, that in the physical interpretation of EFT's, one must take care to distinguish between those aspects of an EFT's formalism that play a central role in generating its successful empirical predictions and those that do not. 
\footnote{This strategy constitutes a central component of what Psillos has labeled the ``divide and conquer" approach to scientific realism \citep{psillos2005scientific}. For a closely related approach to the realist interpretation of quantum theories in general, see Saatsi \citep{saatsi2017scientific}. 
}
As we have seen, it is precisely such considerations that motivate the abandonment of fundamental parameters in the physical interpretation of Wilsonian EFTs, since the notion that there exists a single physically preferred parametrization of the theory at its physical cutoff plays no essential role in generating the theory's successful predictions (or in making it finite). 
\footnote{The situation is partly analogous to the distinction between the mechanical ether theory of Lorentz, which predicted relativistic effects such as time dilation, length contraction and the like, all while assuming a physically preferred frame of reference, and the relativity of Einstein, which recovered the same effects but relinquished the idle assumption of a preferred frame \citep{janssen2002reconsidering}. Until it is possible to independently measure the bare parameters of an EFT (rather than working backwards from correlation functions), the notion of fundamental parameters appears to rest primarily on metaphysical speculation motivated by analogies with condensed matter physics, much as intuitions motivating the introduction of an ether in the 19th century rested in part on metaphysical speculation motivated by analogies with the mechanical theory of elastic media. 
}

However, the precise sense in which Williams and Fraser use the Wilsonian RG  to identify candidates for realist commitment differs from the sense in which the interpretation of Wilsonian EFTs sketched in this section does so. As discussed, Williams and Fraser use the Wilsonian RG to identify elements of an EFT that are robust to changes in the specification of a theory at large values of the parameter $\Lambda$ (typically, near the physical cutoff): since we are ignorant as to the true nature of physics near the physical cutoff (at least, in the case of the Standard Model effective field theory) we should restrict our realism to quantities that do not depend on these details. 

We agree with this claim, and simply wish to note that there is a distinct and compatible sense in which the Wilsonian RG may be used to identify candidates for realist commitment in EFTs. By restricting physical quantities in an EFT to Wilsonian RG invariants, the physical interpretation of EFTs sketched in this section ventures a hypothesis about the \textit{semantic} relationship between an EFT and the physical world, rather than a claim about our \textit{epistemic} relationship to the world. For example, given that there is some matter of fact - albeit currently unknown to us - about the values of the irrelevant operator coefficients that describe physics near the Standard Model's physical cutoff scale, what features of that non-renormalizable EFT's mathematical formalism represent real features in the world, and which are mere artifacts of representation? One possibility is that there really exist fundamental bare parameters that can in principle be directly measured (by analogy with the microscopic parameters of a condensed matter system). Another possibility is that the bare cutoff parameters of that EFT do not represent real features in the world, and instead merely reflect an arbitrary choice of parametrization of physical amplitudes, akin to a choice of gauge or coordinate system. As we have argued, the historical progression of EFTs to ever-higher energies appears to support the latter view: in the case of the non-renormalizable 4-Fermi theory of weak interactions, we possess empirical access to physics near and beyond the physical cutoff, but do not measure the bare cutoff parameters of this EFT directly; there does not appear to be any empirical evidence for the notion that there are genuine physical facts about the values of the bare parameters, even in cases where we can probe physics near the cutoff. Things could be different in the case of Standard Model effective field theory, since we do not yet have access to physics near its cutoff; however, there does not appear to much by way of empirical evidence that this is the case, either. Thus, while the realism of Williams and Fraser appears to be consistent with (if not explicitly to endorse) the physical interpretation of Wilsonian EFTs in terms of fundamental parameters, the interpretation sketched in this section explicitly rejects this possibility on the grounds that these quantities are not invariant under Wilsonian RG transformations.

\section{``Formal" vs. ``Physical" Views of the High-Energy/Condensed-Matter Analogy} \label{FraserAnalogies}

Previously, we saw that one way of understanding the motivation for the naturalness principle, articulated by Schwartz, is through the analogy between high-energy and condensed matter physics \citep{schwartz2014quantum}. In Schwartz' reading, the bare parameters of the Standard Model possess a physical interpretation closely analogous to the microscopic parameters that describe inter-atomic or inter-molecular interactions in a solid state system. This interpretation in turn reinforces the notion that the delicate sensitivity of the Higgs pole mass to these bare parameters constitutes a problematic instance of fine tuning. While it is unclear as a historical and sociological matter just how prevalent this particular reading of Susskind was among practicing particle physicists, from a conceptual point of view, it is a position worth considering. 

From the perspective described in the previous section, in which Wilsonian EFTs do not have fundamental parameters, this particular justification for the naturalness principle appears to rest on an excessively ``literal" or ``physical" interpretation of the analogy between condensed matter physics (CMP) and high energy physics (HEP). To clarify our meaning, it will be useful to employ a distinction between ``formal" and ``physical" interpretations of the high-energy/condensed-matter analogy, which has recently been developed in the work of D. Fraser and of D. Fraser \& A. Koberinski \citep{fraserWilson}, \citep{fraser2016higgs}. 

According to Fraser, a ``formal" analogy occurs when the same mathematical formalism can be applied in physically distinct contexts. For example, one can apply the wave equation to the description of sound waves and electromagnetic waves, which constitute strongly physically distinct domains of application. The formal analogy between high-energy physics and condensed matter physics is evidenced by the fact that in both contexts, one finds the application of path integrals, Feynman diagrams, the renormalization group, and spontaneous symmetry breaking (SSB), to name a few of the main points of structural commonality. 

Within Fraser's view, not all formal analogies are physical analogies. She and Koberinski describe the difference by saying that ``formal analogies map similar elements of the mathematical formalism of the models; physical analogies map elements of the models with similar physical interpretations" \citep{fraser2016higgs}. Naturally, this raises the question of what constitutes similarity of physical interpretation. For our purposes here, it will be sufficient to illustrate the core idea of the distinction by way of example, beginning with a case discussed by Fraser and Koberinski.  

As an example of an analogy that is formal but not physical, Fraser and Koberinski consider the analogy between SSB in the SM Higgs mechanism and in the BCS and Ginzburg-Landau theories of superconductivity. Here, they note that while in condensed matter systems, SSB is a process that can occur through time, in the framework of the Standard Model (in ordinary rather than finite-temperature quantum field theory), the parameters are fixed for all time so that symmetry breaking may not be understood as a temporal process. 
\footnote{On the other hand, their claim may not extend to SSB in the more general framework of finite-temperature quantum field theory. 
}
Thus, while similar mathematical formalisms are employed across the analogy, the physical interpretation of this formalism differs dramatically between the two contexts. 

A second example of an analogy that might be taken as formal but not physical is the analogy between the theory of electromagnetic waves and the theory of mechanical waves (e.g., sound waves). Whereas sound waves propagate in a mechanical medium and their behavior can often be derived on the basis of classical mechanical models of this medium, electromagnetic waves do not arise as disturbances within such a mechanical medium. Nevertheless, the thought that electromagnetic waves might arise as disturbances in such a medium inspired the unsuccessful attempts in the nineteenth century to formulate an ether theory of light. Thus, efforts to formulate such an ether theory could be characterized in Fraser's terms as falsely supposing the analogy between sound waves and electromagnetic waves to be a physical rather than merely formal analogy. 

In regard to the Higgs naturalness problem, the interpretation of bare SM parameters as fundamental parameters in the specific sense articulated by Schwartz may reasonably be characterized as extending the established formal analogy between high energy and condensed matter physics - characterized by the common formalism of path integrals, Feynman diagrams, renormalization group, and SSB - to a physical analogy. It does so by supposing the existence of a physically preferred, ``fundamental" bare parametrization at the SM's physical cutoff, just as in certain condensed matter systems there exists an underlying microscopic lattice description that directly characterizes the microscopic interactions of the atoms in the lattice. The notion that there is a single, true, ``microscopic," fundamental bare parametrization for any given HEP EFT goes beyond recognition of the mathematical similarities between high energy and condensed matter theory to attribute analogous physical interpretations to the bare parameters across both contexts. \citep{fraserWilson} provides a detailed, systematic investigation of the analogy between condensed matter field theory and relativistic quantum field theory in the context of Wilson's work on the renormalization group; she argues that this analogy is merely formal, including with regard to the interpretation of parameters such as the bare mass. 
\footnote{Nevertheless, we wish to highlight as a potential caveat that the formal/physical distinction may not be sharply defined. All pairs of physical systems must be physically disanalogous in \textit{some} sense - otherwise, they would be the same system. Thus, as a qualification to Fraser's distinction, we adopt the view that no analogy is physical \textit{simpliciter}, but rather only physical \textit{with respect to} particular aspects of the analogy, such as (in our case), the status of bare parameters. 
}
	One way of interpreting the analogy as merely formal (though perhaps not Fraser's) is to relinquish the notion that there is a physical matter of fact as to which point along the Wilsonian RG trajectory constitutes the physically correct parametrization of the theory.

\section{Conclusion} \label{Conclusion}

One influential motivation for the naturalness principle is based on the notion of ``fundamental parameters," which are supposed to constitute the physically correct, defining parametrization of an EFT. We have suggested how it might be possible, and desirable, to abandon this notion within the usual Wilsonian, path-integral-based formalism for calculating correlation functions, without sacrificing predictive power or the virtue of finiteness introduced by the Wilsonian approach. This alternative view, in which a Wilsonian EFT does not have a single physically preferred parametrization, and is defined instead by its entire RG trajectory, is motivated primarily by the fact that a continuous infinity of different parametrizations serve equally well to generate the EFT's successful predictions through its correlation functions, and that the notion of a single fundamental parametrization may therefore be unnecessary to the formulation of the theory. We have argued that this way of understanding Wilsonian EFTs also invites abandonment of several other traditional dogmas of Wilsonian effective field theory, such as the notion that an EFT should be mathematically defined with a cutoff equal to its physical cutoff. However, we emphasize that while the choice of a single preferred bare parametrization is unnecessary to generating the successful predictions of an EFT from its correlation functions, the question of how to specify the Hilbert space and Hamiltonian of such a theory without relying on a fundamental parametrization remains unanswered. 

Within the alternative understanding of Wilsonian EFTs sketched here, it is possible to understand the delicate cancellation between Higgs bare mass and quantum corrections as an eliminable artifact resulting from a particular choice of mathematical convention, and the problem of explaining these cancellations seems much less urgent, if not wholly misguided. Our discussion of this view has been an attempt to explicitly draw out the implications for the formulation and interpretation of quantum field theory of Wetterich's claim that Higgs fine tuning is merely an artifact of a bad expansion method. Moreover, we emphasize that this way of understanding Wilsonian EFTs without fundamental parameters does nothing to undermine the expectation that the Standard Model as a whole should be embedded in a deeper theory that includes gravity (which should explain why Standard Model effective field theory is characterized by the particular RG trajectory that it is, rather than the value of any specific point on it). Rather, it only weakens the case that there is an especially urgent need for explanation attached to the Higgs mass and not other SM parameters.

There also exist other formulations of the naturalness principle that we have not considered here, such as fine tuning formulations in terms of renormalized parameters and formulations based on the notion of inter-scale autonomy, each of which needs to be examined in detail. We think it likely that similar concerns to those that arise in the context of Higgs bare mass fine tuning will arise in the context of these other formulations, as well as in the cosmological constant problem. However, in each of these instances, a rigorous case remains to be made. 

\

\noindent \textbf{Acknowledgments:} This work was supported by the DFG, grant FOR 2063. We would like to thank the members of the DFG Research Unit ``The Epistemology of the Large Hadron Collider," in particular M.A. Carretero Sahuquillo, R. Dardashti, and G. Schiemann for fruitful discussions. We would also like to thank Doreen Fraser for helpful comments on an earlier draft of this article.

\bibliographystyle{plainnat}
\bibliography{References.bib}

\begin{thebibliography}{67}
\providecommand{\natexlab}[1]{#1}
\providecommand{\url}[1]{\texttt{#1}}
\expandafter\ifx\csname urlstyle\endcsname\relax
  \providecommand{\doi}[1]{doi: #1}\else
  \providecommand{\doi}{doi: \begingroup \urlstyle{rm}\Url}\fi

\bibitem[Anderson and Castano(1995)]{anderson1995measures}
Greg Anderson and Diego Castano.
\newblock Measures of fine tuning.
\newblock \emph{Physics Letters B}, 347\penalty0 (3-4):\penalty0 300--308,
  1995.

\bibitem[Bain(2013)]{bain2013emergence}
Jonathan Bain.
\newblock Emergence in effective field theories.
\newblock \emph{European journal for philosophy of science}, 3\penalty0
  (3):\penalty0 257--273, 2013.

\bibitem[Barbieri and Giudice(1988)]{barbieri1988upper}
Riccardo Barbieri and Gian~Francesco Giudice.
\newblock Upper bounds on supersymmetric particle masses.
\newblock \emph{Nuclear Physics B}, 306\penalty0 (1):\penalty0 63--76, 1988.

\bibitem[Berges et~al.(2002)Berges, Tetradis, and Wetterich]{berges2002non}
J{\"u}rgen Berges, Nikolaos Tetradis, and Christof Wetterich.
\newblock Non-perturbative renormalization flow in quantum field theory and
  statistical physics.
\newblock \emph{Physics Reports}, 363\penalty0 (4):\penalty0 223--386, 2002.

\bibitem[Bianchi and Rovelli(2010)]{bianchi2010all}
Eugenio Bianchi and Carlo Rovelli.
\newblock Why all these prejudices against a constant?
\newblock \emph{arXiv:1002.3966}, 2010.

\bibitem[Butterfield(2011{\natexlab{a}})]{butterfield2011emergence}
Jeremy Butterfield.
\newblock Emergence, reduction and supervenience: a varied landscape.
\newblock \emph{Foundations of Physics}, 41\penalty0 (6):\penalty0 920--959,
  2011{\natexlab{a}}.

\bibitem[Butterfield(2011{\natexlab{b}})]{butterfield2011less}
Jeremy Butterfield.
\newblock Less is different: emergence and reduction reconciled.
\newblock \emph{Foundations of Physics}, 41\penalty0 (6):\penalty0 1065--1135,
  2011{\natexlab{b}}.

\bibitem[Butterfield(2014)]{butterfield2014reduction}
Jeremy Butterfield.
\newblock Reduction, emergence, and renormalization.
\newblock \emph{The Journal of Philosophy}, 111\penalty0 (1):\penalty0 5--49,
  2014.

\bibitem[Butterfield and Bouatta(2012)]{butterfield2012emergence}
Jeremy Butterfield and Nazim Bouatta.
\newblock Emergence and reduction combined in phase transitions.
\newblock In \emph{Frontiers of Fundamental Physics: The Eleventh International
  Symposium}, volume 1446, pages 383--403. AIP Publishing, 2012.

\bibitem[Butterfield and Bouatta(2015)]{butterfield2015renormalization}
Jeremy Butterfield and Nazim Bouatta.
\newblock Renormalization for philosophers.
\newblock \emph{Metaphysics in Contemporary Physics}, 104:\penalty0 437--485,
  2015.

\bibitem[Castellani(2002)]{castellani2002reductionism}
Elena Castellani.
\newblock Reductionism, emergence, and effective field theories.
\newblock \emph{Studies in History and Philosophy of Science Part B: Studies in
  History and Philosophy of Modern Physics}, 33\penalty0 (2):\penalty0
  251--267, 2002.

\bibitem[Collins(1984)]{collins1984renormalization}
John Collins.
\newblock \emph{Renormalization: an introduction to renormalization, the
  renormalization group and the operator-product expansion}.
\newblock Cambridge university press, 1984.

\bibitem[Crowther(2015)]{crowther2015decoupling}
Karen Crowther.
\newblock Decoupling emergence and reduction in physics.
\newblock \emph{European Journal for Philosophy of Science}, 5\penalty0
  (3):\penalty0 419--445, 2015.

\bibitem[Crowther(2017)]{crowther2017inter}
Karen Crowther.
\newblock Inter-theory relations in quantum gravity: Correspondence, reduction,
  and emergence.
\newblock \emph{arXiv:1712.00473}, 2017.

\bibitem[Delamotte(2012)]{delamotte2012introduction}
Bertrand Delamotte.
\newblock An introduction to the nonperturbative renormalization group.
\newblock In \emph{Renormalization Group and Effective Field Theory Approaches
  to Many-Body Systems}, pages 49--132. Springer, 2012.

\bibitem[Fletcher(2015)]{fletcher2015similarity}
Samuel Fletcher.
\newblock Similarity, topology, and physical significance in relativity theory.
\newblock \emph{The British Journal for the Philosophy of Science}, 67\penalty0
  (2):\penalty0 365--389, 2015.

\bibitem[Franklin(2017)]{franklin2017whence}
Alexander Franklin.
\newblock Whence the effectiveness of effective field theories?
\newblock \emph{The British Journal for the Philosophy of Science}, DOI:
  \url{https://doi.org/10.1093/bjps/axy050}, 2017.

\bibitem[Fraser(2009)]{fraser2009quantum}
Doreen Fraser.
\newblock Quantum field theory: Underdetermination, inconsistency, and
  idealization.
\newblock \emph{Philosophy of Science}, 76\penalty0 (4):\penalty0 536--567,
  2009.

\bibitem[Fraser(2011)]{fraser2011take}
Doreen Fraser.
\newblock How to take particle physics seriously: A further defence of
  axiomatic quantum field theory.
\newblock \emph{Studies In History and Philosophy of Science Part B: Studies In
  History and Philosophy of Modern Physics}, 42\penalty0 (2):\penalty0
  126--135, 2011.

\bibitem[Fraser(2018)]{fraserWilson}
Doreen Fraser.
\newblock The development of renormalization group methods for particle
  physics: Formal analogies between classical statistical mechanics and quantum
  field theory.
\newblock \emph{Preprint}, 2018.
\newblock URL \url{\url{http://philsci-archive.pitt.edu/14591/}}.

\bibitem[Fraser and Koberinski(2016)]{fraser2016higgs}
Doreen Fraser and Adam Koberinski.
\newblock The {H}iggs mechanism and superconductivity: A case study of formal
  analogies.
\newblock \emph{Studies in History and Philosophy of Science Part B: Studies in
  History and Philosophy of Modern Physics}, 55:\penalty0 72--91, 2016.

\bibitem[Fraser(2016)]{fraser2016quantum}
James Fraser.
\newblock \emph{What is Quantum Field Theory? Idealisation, Explanation and
  Realism in High Energy Physics.}
\newblock PhD thesis, University of Leeds, 2016.

\bibitem[Fraser(2017)]{fraser2017real}
James Fraser.
\newblock The real problem with perturbative quantum field theory.
\newblock \emph{The British Journal for the Philosophy of Science}, DOI:
  \url{https://doi.org/10.1093/bjps/axx042}, 2017.

\bibitem[Giudice(2013)]{giudice2013naturalness}
Gian Giudice.
\newblock Naturalness after {LHC8}.
\newblock \emph{arXiv:1307.7879}, 2013.

\bibitem[Giudice(2017)]{giudice2017dawn}
Gian Giudice.
\newblock The dawn of the post-naturalness era.
\newblock \emph{arXiv:1710.07663}, 2017.

\bibitem[Halvorson(2013)]{HalvorsonSynSem}
Hans Halvorson.
\newblock The semantic view, if plausible, is syntactic.
\newblock \emph{Philosophy of Science}, 80\penalty0 (3), 2013.

\bibitem[Hancox-Li(2017)]{hancox2017solutions}
Leif Hancox-Li.
\newblock Solutions in constructive field theory.
\newblock \emph{Philosophy of Science}, 84\penalty0 (2):\penalty0 335--358,
  2017.

\bibitem[Hartmann(2001)]{hartmann2001effective}
Stephan Hartmann.
\newblock Effective field theories, reductionism and scientific explanation.
\newblock \emph{Studies in History and Philosophy of Science Part B: Studies in
  History and Philosophy of Modern Physics}, 32\penalty0 (2):\penalty0
  267--304, 2001.

\bibitem[Hossenfelder(2018)]{hossenfelder2018screams}
Sabine Hossenfelder.
\newblock Screams for explanation: Finetuning and naturalness in the
  foundations of physics.
\newblock \emph{arXiv:1801.02176}, 2018.

\bibitem[Huggett and Weingard(1995)]{huggett1995renormalisation}
Nick Huggett and Robert Weingard.
\newblock The renormalisation group and effective field theories.
\newblock \emph{Synthese}, 102\penalty0 (1):\penalty0 171--194, 1995.

\bibitem[Janssen(2002)]{janssen2002reconsidering}
Michel Janssen.
\newblock Reconsidering a scientific revolution: The case of {E}instein versus
  {L}orentz.
\newblock \emph{Physics in perspective}, 4\penalty0 (4):\penalty0 421--446,
  2002.

\bibitem[Martin(2010)]{martin2010supersymmetry}
Stephen Martin.
\newblock A supersymmetry primer.
\newblock In \emph{Perspectives on supersymmetry II}, pages 1--153. World
  Scientific, 2010.

\bibitem[Miller(2016)]{miller2016mathematical}
Michael Miller.
\newblock Mathematical structure and empirical content.
\newblock 2016.
\newblock URL \url{\url{http://philsci-archive.pitt.edu/12678/}}.

\bibitem[Montvay and M{\"u}nster(1997)]{montvay1997quantum}
Istv{\'a}n Montvay and Gernot M{\"u}nster.
\newblock \emph{Quantum fields on a lattice}.
\newblock Cambridge University Press, 1997.

\bibitem[Morris(1994)]{morris1994exact}
Tim Morris.
\newblock The exact renormalization group and approximate solutions.
\newblock \emph{International Journal of Modern Physics A}, 9\penalty0
  (14):\penalty0 2411--2449, 1994.

\bibitem[Morris(1998)]{morris1998elements}
Tim Morris.
\newblock Elements of the continuous renormalization group.
\newblock \emph{Progress of Theoretical Physics Supplement}, 131:\penalty0
  395--414, 1998.

\bibitem[Nelson and Riedel(2017)]{nelson2017classical}
Elliot Nelson and C.~Jess Riedel.
\newblock Classical branches and entanglement structure in the wavefunction of
  cosmological fluctuations.
\newblock \emph{arXiv:1711.05719}, 2017.

\bibitem[Peskin and Schroeder(1995)]{peskin1995introduction}
Michael Peskin and Daniel Schroeder.
\newblock \emph{An Introduction To Quantum Field Theory (Frontiers in
  Physics)}.
\newblock Westview Press Incorporated, 1995.

\bibitem[Polchinski(1984)]{polchinski1984renormalization}
Joseph Polchinski.
\newblock Renormalization and effective {L}agrangians.
\newblock \emph{Nuclear Physics B}, 231\penalty0 (2):\penalty0 269--295, 1984.

\bibitem[Psillos(2005)]{psillos2005scientific}
Stathis Psillos.
\newblock \emph{Scientific realism: How science tracks truth}.
\newblock Routledge, 2005.

\bibitem[Psillos(2011)]{psillos2011moving}
Stathis Psillos.
\newblock Moving molecules above the scientific horizon: On {P}errin's case for
  realism.
\newblock \emph{Journal for General Philosophy of Science}, 42\penalty0
  (2):\penalty0 339--363, 2011.

\bibitem[Ranard(2015)]{ranard2015introduction}
Daniel Ranard.
\newblock An introduction to rigorous formulations of quantum field theory.
\newblock Unpublished essay, advisors: Ryszard Kostecki and Lucien Hardy,
  Perimeter Institute for Theoretical Physics, 2015.

\bibitem[Rivasseau(2014)]{rivasseau2014perturbative}
Vincent Rivasseau.
\newblock \emph{From perturbative to constructive renormalization}.
\newblock Princeton University Press, 2014.

\bibitem[Rosaler(2013)]{RosalerThesis}
Joshua Rosaler.
\newblock \emph{Inter-theory Relations in Physics: Case Studies from Quantum
  Mechanics and Quantum Field Theory}.
\newblock PhD thesis, University of Oxford, 2013.

\bibitem[Rosaler(2015)]{rosaler2015local}
Joshua Rosaler.
\newblock Local reduction in physics.
\newblock \emph{Studies in History and Philosophy of Science Part B: Studies in
  History and Philosophy of Modern Physics}, 50:\penalty0 54--69, 2015.

\bibitem[Rosaler(2017)]{rosaler2017reduction}
Joshua Rosaler.
\newblock Reduction as an a posteriori relation.
\newblock \emph{The British Journal for the Philosophy of Science}, DOI:
  \url{https://doi.org/10.1093/bjps/axx026}, 2017.

\bibitem[Rosaler(2018)]{rosaler2018generalized}
Joshua Rosaler.
\newblock Generalized {E}hrenfest relations, deformation quantization, and the
  geometry of inter-model reduction.
\newblock \emph{Foundations of Physics}, 48\penalty0 (3):\penalty0 355--385,
  2018.

\bibitem[Rosten(2012)]{rosten2012fundamentals}
Oliver Rosten.
\newblock Fundamentals of the exact renormalization group.
\newblock \emph{Physics Reports}, 511\penalty0 (4):\penalty0 177--272, 2012.

\bibitem[Ruetsche(2011)]{ruetsche2011interpreting}
Laura Ruetsche.
\newblock \emph{Interpreting quantum theories}.
\newblock Oxford University Press, 2011.

\bibitem[Ruetsche(2018)]{ruetscheAscent}
Laura Ruetsche.
\newblock Renormalization group realism: The ascent of pessimism.
\newblock \emph{Philosophy of Science}, 85\penalty0 (5):\penalty0 1176--1189,
  2018.

\bibitem[Saatsi(2017)]{saatsi2017scientific}
Juha Saatsi.
\newblock Scientific realism meets metaphysics of quantum mechanics.
\newblock 2017.
\newblock URL \url{\url{http://philsci-archive.pitt.edu/14583/}}.

\bibitem[Schwartz(2014)]{schwartz2014quantum}
Matthew Schwartz.
\newblock \emph{Quantum field theory and the standard model}.
\newblock Cambridge University Press, 2014.

\bibitem[Skiba(2010)]{skiba2010tasi}
Witold Skiba.
\newblock {TASI} lectures on effective field theory and precision electroweak
  measurements.
\newblock \emph{arXiv:1006.2142}, 2010.

\bibitem[Srednicki(2007)]{srednicki2007quantum}
Mark Srednicki.
\newblock \emph{Quantum field theory}.
\newblock Cambridge University Press, 2007.

\bibitem[Susskind(1979)]{susskind1979dynamics}
Leonard Susskind.
\newblock Dynamics of spontaneous symmetry breaking in the {S}alam-{Weinberg}
  theory.
\newblock \emph{Physical Review D}, 20\penalty0 (10):\penalty0 2619, 1979.

\bibitem['t~Hooft(1980)]{hooft1980naturalness}
Gerard 't~Hooft.
\newblock Naturalness, chiral symmetry, and spontaneous chiral symmetry
  breaking.
\newblock \emph{Recent developments in gauge theories. Proceedings of the NATO
  Advanced Study Institute on recent developments in gauge theories, held in
  Carg{\`e}se, Corsica, August 26-September 8, 1979}, pages 135--157, 1980.

\bibitem[Wallace(2006)]{wallaceNaivete}
David Wallace.
\newblock In defence of naivete: The conceptual status of {L}agrangian quantum
  field theory.
\newblock \emph{Synthese}, 151:\penalty0 33--80, 2006.

\bibitem[Wallace(2011)]{wallace2011taking}
David Wallace.
\newblock Taking particle physics seriously: A critique of the algebraic
  approach to quantum field theory.
\newblock \emph{Studies in History and Philosophy of Science Part B: Studies in
  History and Philosophy of Modern Physics}, 42\penalty0 (2):\penalty0
  116--125, 2011.

\bibitem[Wells(2015)]{wells2015utility}
James Wells.
\newblock The utility of naturalness, and how its application to quantum
  electrodynamics envisages the standard model and {H}iggs boson.
\newblock \emph{Studies in History and Philosophy of Science Part B: Studies in
  History and Philosophy of Modern Physics}, 49:\penalty0 102--108, 2015.

\bibitem[Wetterich(1984)]{wetterich1984fine}
Christof Wetterich.
\newblock Fine-tuning problem and the renormalization group.
\newblock \emph{Physics Letters B}, 140\penalty0 (3-4):\penalty0 215--222,
  1984.

\bibitem[Wetterich(1991)]{wetterich1991average}
Christof Wetterich.
\newblock Average action and the renormalization group equations.
\newblock \emph{Nuclear Physics B}, 352\penalty0 (3):\penalty0 529--584, 1991.

\bibitem[Wetterich(1993)]{wetterich1993exact}
Christof Wetterich.
\newblock Exact evolution equation for the effective potential.
\newblock \emph{Physics Letters B}, 301\penalty0 (1):\penalty0 90--94, 1993.

\bibitem[Wetterich(2012)]{wetterich2012look}
Christof Wetterich.
\newblock Where to look for solving the gauge hierarchy problem?
\newblock \emph{Physics Letters B}, 718\penalty0 (2):\penalty0 573--576, 2012.

\bibitem[Williams(2015)]{williams2015naturalness}
Porter Williams.
\newblock Naturalness, the autonomy of scales, and the 125 {GeV} {H}iggs.
\newblock \emph{Studies in History and Philosophy of Science Part B: Studies in
  History and Philosophy of Modern Physics}, 51:\penalty0 82--96, 2015.

\bibitem[Williams(2017)]{williams2017scientific}
Porter Williams.
\newblock Scientific realism made effective.
\newblock \emph{The British Journal for the Philosophy of Science}, DOI:
  \url{https://doi.org/10.1093/bjps/axx043}, 2017.

\bibitem[Wimsatt(2007)]{wimsatt2007re}
William Wimsatt.
\newblock \emph{Re-engineering philosophy for limited beings: Piecewise
  approximations to reality}.
\newblock Harvard University Press, 2007.

\bibitem[Woit(2014)]{woitEdge}
Peter Woit.
\newblock What scientific idea is ready for retirement? {T}he ``naturalness"
  argument.
\newblock \emph{Edge}, 2014.
\newblock URL \url{\url{https://www.edge.org/response-detail/25416}}.

\end{thebibliography}
\end{document}